\documentclass[twocolumn,twocolappendix]{aastex7}
\usepackage{booktabs}
\usepackage{CJK}

\defcitealias{Kerr21}{SPYGLASS-I}
\defcitealias{Kerr22b}{SPYGLASS-III}
\defcitealias{Kerr22a}{SPYGLASS-II}
\defcitealias{Kerr23}{SPYGLASS-IV}
\defcitealias{Kerr24}{SPYGLASS-V}
\defcitealias{Kerr25}{SPYGLASS-VI}
\defcitealias{Kerr25b}{SPYGLASS-VII-A}

\begin{document}

\begin{CJK*}{UTF8}{gbsn}

\title{SPYGLASS. VII-B. Tracing the Fragments of Massive Star Formation Using Low-Mass Associations}

\author[0000-0002-6549-9792]{Ronan Kerr}
\affiliation{Dunlap Institute for Astronomy \& Astrophysics, University of Toronto,
Toronto, ON M5S 3H4, Canada\\}
\affiliation{Department of Astronomy, University of Texas at Austin, 2515 Speedway, Stop C1400, Austin, Texas, USA 78712-1205\\}
\email{ronan.kerr@utoronto.ca}

\author[0000-0001-9811-568X]{Adam L. Kraus}
\affiliation{Department of Astronomy, University of Texas at Austin, 2515 Speedway, Stop C1400, Austin, Texas, USA 78712-1205\\}
\email{alk@astro.as.utexas.edu}

\author[0000-0002-3389-9142]{Jonathan C. Tan}
\affiliation{Dept. of Space, Earth \& Environment, Chalmers University of Technology, Gothenburg, Sweden}
\affiliation{Dept. of Astronomy \& Virginia Institute for Theoretical Astronomy, University of Virginia, Charlottesville, VA, USA}
\email{jctan.astro@gmail.com}

\author[0000-0003-2481-4546]{Julio Chanam\'e}
\affiliation{Instituto de Astrof\'isica, Pontificia Universidad Cat\'olica de Chile, Avenida Vicu\~na Mackenna 4860, 782-0436 Macul, Santiago, Chile}
\email{jchaname@astro.puc.cl}

\author[0000-0002-4128-7867]{Facundo P\'erez Paolino}
\affiliation{Department of Astronomy, California Institute of Technology, 1216 East California Blvd, Pasadena, CA 91125, USA\\}
\email{fperezpa@caltech.edu}

\author[0000-0003-2573-9832]{Joshua S. Speagle(沈佳士\ignorespacesafterend)}
\affiliation{Dunlap Institute for Astronomy \& Astrophysics, University of Toronto,
Toronto, ON M5S 3H4, Canada\\}
\affiliation{Department of Statistical Sciences, University of Toronto\\ 9th Floor, Ontario Power Building, 700 University Ave, Toronto, ON M5G 1Z5, Canada}
\affiliation{David A. Dunlap Department of Astronomy \& Astrophysics, University of Toronto\\ 50 St George Street, Toronto, ON M5S 3H4, Canada}
\affiliation{Data Sciences Institute, University of Toronto\\ 17th Floor, Ontario Power Building, 700 University Ave, Toronto, ON M5G 1Z5, Canada} 
\email{j.speagle@utoronto.ca}

\author[0000-0002-5851-2602]{Juan P. Farias}
\affiliation{Department of Physics and Astronomy, McMaster University, 1280 Main Street West, Hamilton, ON, L8S 4M1, Canada\\}
\email{fariasoj@mcmaster.ca}

\author[0000-0003-3526-5052]{Jos\'e G. Fern\'andez-Trincado}
\affiliation{Universidad Cat\'olica del Norte, N\'ucleo UCN en Arqueolog\'ia Gal\'actica, Av. Angamos 0610, Antofagasta, Chile}
\affiliation{Universidad Cat\'olica del Norte, Departamento de Ingenier\'ia de Sistemas y Computaci\'on, Av. Angamos 0610, Antofagasta, Chile}
\email{jose.fernandez@ucn.cl}

\author[0000-0002-1423-2174]{Keith Hawkins}
\affiliation{Department of Astronomy, University of Texas at Austin, 2515 Speedway, Stop C1400, Austin, Texas, USA 78712-1205\\}
\email{keithhawkins@utexas.edu}

\begin{abstract}

New observations from the \textit{Gaia} spacecraft have traced an emerging demographic of low-mass associations disconnected from larger associations or GMCs. The first of these associations were recently characterized, but the star-forming environments they trace remain unknown. Using new velocities and ages alongside literature catalogs, we uncover the origins of 16 low-mass associations ($M\lesssim100$ M$_{\odot}$, $\tau\lesssim50$ Myr) using dynamical traceback. We reveal that three groups of currently disparate populations share common formation sites, comprising the Leo, CaNMoS, and AquENS associations. Twelve of 16 associations have plausible connections to larger complexes, six of which form while moving outward from well-established multi-generational star-forming events that drive known or suspected bubbles. We find that feedback from the oldest co-spatial and co-moving relatives of these associations can explain the current morphologies of the Local and Orion-Eridanus Bubbles, along with the formation of related associations like Sco-Cen and Orion OB1. Most remaining populations show evidence for triggered star formation. In the Leo Association, high vertical velocities and a deceleration signature suggest that it formed out of an intermediate velocity cloud colliding with gas in Orion, which would make it the first known case of star formation in one of these clouds. The other newly defined associations show similar asymmetric velocity signatures, such as CaNMoS, which may trace bubble-driven acceleration or a cloud collision. We conclude that the lowest-mass young associations remain undiscovered, and that these populations may have a critical role revealing the small gas overdensities that trace the processes sculpting galactic star formation.

\end{abstract}



\section{Introduction}

Star formation in the solar neighborhood typically produces stellar associations, unbound groups of co-moving stars that disperse after formation \citep{deZeeuw99, lada03}. These associations provide a fossil record of the gas clouds that precede them, retaining the motions of their parent cloud and allowing sites of star formation to be reconstructed tens of Myr after gas dispersal \citep{Gagne21, Kerr22a, Swiggum24}. These young associations are therefore essential to our understanding of how star formation evolves in our Galaxy, providing a time-resolved view that is otherwise inaccessible. Many large associations in the solar neighborhood like Sco-Cen, Cep-Her, and Vela have had their star formation histories studied extensively, revealing evidence for feedback-driven star formation and other mechanisms by which star formation can propagate from one event to the next \citep[e.g.,][]{Pang21, Kerr24, Posch25}. However, many smaller young associations have only recently been recognized, and their role within the local star formation record remains unknown \citep{Kerr23}. 

The \textit{Gaia} spacecraft \citep{GaiaMission, GaiaDR322} has dramatically expanded our knowledge of the local star formation record, enabling surveys that have discovered hundreds of new young stellar populations \citep{Kounkel19, Kerr21, Prisinzano22, Hunt23}. The SPYGLASS program, which uses \textit{Gaia} photometry and astrometry to identify probable young stars ($\tau\lesssim50$~Myr) and group them into associations, has produced particularly sensitive surveys, with \citetalias{Kerr23} revealing 10 entirely new groups and 20 with no direct literature equivalents. Most new discoveries are part of an emerging demographic of low-mass stellar populations ($M\lesssim100$~M$_{\odot}$) that has been largely unstudied to date. \citet{Kerr25b} (hereafter SPYGLASS-VII-A) computed  bulk positions, velocities, and demographic properties for 15 of these low-mass associations for the first time, in addition to ages from Lithium depletion, isochrones, and dynamics. That work also identified substructure, revealing that Taurus-Orion 1 consists of two components, TOR1A and TOR1B, with motions inconsistent with a common origin. The new TOR1B association and three others in the set have total masses under 30 M$_{\odot}$, making them among the smallest associations ever characterized. 

The conditions under which low-mass associations like these form are unclear. Unlike massive star-forming events like Orion and Sco-Cen \citep{Kerr23}, low-mass associations require far smaller parent clouds, which are distributed throughout the local volume \citep{Cahlon24}. They may therefore trace overdensities within far finer galactic structure, such as the edges of expanding bubbles or sparser sections of spiral arm-driven gas ridges \citep{Zucker22b}. Several low-mass associations have been seen on the edges of larger complexes, such as TW Hydrae and $\eta$ Cha in Sco-Cen \citep{Kastner97, Kerr21, MiretRoig25}. It may therefore follow that low-mass associations originate in gas fragments accelerated out of larger complexes, or out of bubbles driven by feedback from those larger star-forming events \citep{Zucker22b}. Many other origins are possible, however, from galactic arm, spur or feather structures \citep[e.g., see][]{Dobbs06}, to molecular cloud collisions \citep[e.g.,][]{Tan00, Wu17}, to collisions with infalling halo gas, which may accompany distinct metallicities \citep[e.g.,][]{Rohser16}. These young associations may provide a record of these processes, indicating the role each plays in local star formation.

In this paper, we place a newly-characterized set of low-mass associations within the context of local star formation, and investigate whether these low-mass associations have plausible connections to larger structures. In Section \ref{sec:data}, we introduce the datasets used to uncover the origins of these populations, and discuss our methods for this analysis in Section \ref{sec:methods}. We then perform our main analysis in Section \ref{sec:results}, where we use dynamical traceback to connect the low-mass associations in our sample to larger structures. We assess the coherence of those patterns and discuss their implications in Section \ref{sec:discussion}, before concluding in Section \ref{sec:conclusion}. 

\section{Data} \label{sec:data}

\subsection{Low-Mass Association Sample}

Our main dataset comes from \citetalias{Kerr25b}, which assessed the ages, membership, and stellar demographics for 16 nearby low-mass associations: Andromeda South (AndS), Aquila East (AqE), Aries South (AriS), Canis Major North (CMaN), Cassiopeia East (CasE), Leo Central (LeoC), Leo East (LeoE), Ophiuchus Southeast (OphSE), Scutum North (ScuN), Theia 72, Theia 78, Vulpecula East (VulE), SCYA-54, SCYA-79, TOR1A, and TOR1B. These populations were drawn from \citetalias{Kerr23}, which reveals 116 young populations and complexes including both well-known star-forming complexes and the new, low-mass populations that form the basis of \citetalias{Kerr25b}'s sample. For each low-mass association, \citetalias{Kerr25b} lists bulk properties, as well as ages that can be used together to reconstruct their sites of formation. 

The stellar catalog from \citetalias{Kerr25b} contains 4244 candidate members distributed across all 16 associations (69 to 817 per association). For each star, \citetalias{Kerr25b} provides \textit{Gaia} astrometry and photometry as well as the membership probabilities from photometric youth and space-transverse velocity agreement with other members. It also provides spectroscopic data from both literature sources and new observations that provide radial velocities (RVs). This dataset therefore provides robust 6D space-velocity vectors for many candidate members, in addition to robust membership assessments. 

\subsection{Literature Sample}

To assess connections between our low-mass young associations and larger star-forming events, we assemble a sample of more massive clusters and literature associations against which to compare. We first select the most numerous populations in the Montreal Open Clusters and Associations database (MOCA; \citealt{Gagne26})\footnote{publicly available at https://mocadb.ca}, considering structures at the lowest level of the hierarchical tree within 500 pc with at least 250 candidate members and catalog ages younger than 80 Myr. MOCA is an inhomogeneous catalog in both age and membership definitions, so additional vetting is required. We remove populations with membership defined solely by \citetalias{Kerr23}, as this work has more generous criteria for the inclusion of candidate members compared to other sources. We further vet the sample by removing populations that are fully undetected in \citetalias{Kerr23}. Several young associations are known only through \citetalias{Kerr23}, including one with a $\sim$30 Myr age and $d\sim400$ pc, so populations excluded from SPYGLASS are unlikely to be both in a relevant age bracket and massive enough to substantially impact nearby star formation. We do not apply this cut to well-characterized populations within 100 pc like the Austral complex, as the use of 2D velocities in SPYGLASS makes it insensitive to populations within this radius.

In the resulting sample, Sco-Cen, Perseus-Orion, Cep-Her, Vela, the Austral Complex, Alessi 20, $\alpha$ Per, and Trumpler 10 all have multiple representatives. Not all SPYGLASS structures are dynamically homogeneous, so for each substructured population, we simplify the sample by selecting individual groups within them to represent major substructures. In Cep-Her, for example, \citetalias{Kerr24} identified 4 major substructures, 3 of which are younger than 80 Myr old: Cinyras, Orpheus, and Cupavo. While all three have internal dynamical variation, all show expansion signatures consistent with their ages, so a single well-populated central cluster can represent their central motion. Outlying structures may have an important role filling in the star formation record, but for connecting our low-mass associations to larger structures, using a single representative is both sufficient and simpler to interpret. In Cep-Her, we select RSG 5 as the representative of Cinyras, $\delta$ Lyr for Orpheus, and NGC 7058 for Cupavo. We follow a similar approach in the other substructured associations, taking more than one representative in populations with multiple well-known massive open clusters like \citet{CantatGaudin19} subgroup IV in Vela. Finally, we add representatives of two notable ($\sim500$ member) complexes missed in our initial selection due to a lack of large central clusters: \citetalias{Kerr22a} subgroup 1 in Cepheus Far North, and L1524 in the Taurus Molecular Cloud \citep{Luhman22}. We list the 45 populations selected for traceback in Table \ref{tab:externalpops}. 

\begin{deluxetable*}{ccccccccccccccccccc}
\tablecolumns{19}
\tablewidth{0pt}
\tabletypesize{\scriptsize}
\tablecaption{Properties of literature populations used in traceback. Most values are from the indicated literature sources, but for populations with inconsistent literature RVs, we recompute RVs using new APOGEE measurements.}
\label{tab:externalpops}
\tablehead{
\colhead{CPLX\tablenotemark{a}} &
\colhead{NAME} &
\multicolumn{2}{c}{Citations\tablenotemark{b}} &
\colhead{Age} &
\colhead{$d$} &
\colhead{$v_r$} &
\colhead{$\mu_{RA}$} &
\colhead{$\mu_{Dec}$} &
\colhead{RA} &
\colhead{Dec} &
\colhead{X} &  
\colhead{Y} &  
\colhead{Z} &  
\colhead{U} &  
\colhead{V} &  
\colhead{W} \\
\colhead{} &
\colhead{} &
\colhead{age} &
\colhead{dynam.} &
\colhead{(Myr)} &
\colhead{(pc)} &
\colhead{(km/s)} &
\multicolumn{2}{c}{(mas yr$^{-1}$)} &
\multicolumn{2}{c}{(deg)} &
\multicolumn{3}{c}{(pc)}&
\multicolumn{3}{c}{(km/s)}}
\startdata
\toprule
AL20 & Alessi20 & 8 & 9/25 & 9.3 & 427 & -10.6 & 8.2 & -2.3 & 2.6 & 58.7 & -198 & 378 & -28 & -8.7 & -17.1 & -6.5 \\
AL20 & RSG7 & 8 & 9/25& 38.9 & 423 & -11.3 & 4.9 & -1.9 & 344.2 & 59.4 & -136 & 400 & -2 & -3.3 & -13.1 & -7.5 \\
 & $\alpha$Per & 17 & 9/15 & 79.0 & 176 & 0.7 & 22.9 & -25.4 & 51.6 & 49.0 & -147 & 94 & -20 & -14.8 & -23.4 & -6.7 \\
AUS & $\chi^1$For & 20 & 20 & 32.4 & 104 & 17.7 & 36.4 & -4.6 & 52.0 & -35.6 & -32 & -49 & -86 & -12.4 & -21.7 & -4.3 \\
AUS & THA & 24 & 20 & 45.7 & 46 & 11.6 & 91.1 & -15.1 & 36.5 & -56.8 & 4 & -25 & -38 & -9.6 & -21.0 & -1.0 \\
CFL & ASCC127 & 8 & 9/15 & 18.2 & 376 & -13.8 & 7.5 & -1.7 & 347.2 & 65.0 & -142 & 346 & 28 & -5.2 & -16.4 & -8.9 \\
SC-BR & IC2602 & 4 & 3/15 & 43.7 & 153 & 17.6 & -17.8 & 10.7 & 160.6 & -64.1 & 51 & -143 & -12 & -8.2 & -21.6 & -0.7 \\
SC-BR & Pl8 & 12 & 12/A  & 37.0 & 138 & 23.2 & -16.0 & 14.0 & 138.3 & -57.7 & 17 & -136 & -15 & -10.8 & -24.4 & -3.8 \\
SC-LC & $\sigma$Cen & 23 & 22/M & 15.5 & 115 & 12.0 & -33.2 & -13.7 & 190.5 & -54.5 & 60 & -96 & 17 & -8.3 & -20.5 & -6.2 \\
SC-US & $\delta$Sco & 23 & 22/M  & 9.8 & 142 & -5.7 & -12.2 & -23.9 & 240.4 & -22.7 & 130 & -22 & 54 & -6.1 & -16.5 & -7.2 \\
SC-UC & eLup & 23 & 22/M  & 20.9 & 145 & 5.6 & -20.8 & -21.7 & 227.7 & -44.7 & 120 & -77 & 29 & -5.1 & -20.4 & -4.0 \\
SC-CR & CrA & 23 & 22/M & 8.5 & 155 & 0.1 & 4.6 & -27.1 & 285.2 & -36.9 & 148 & 0 & -47 & -3.1 & -17.4 & -9.8  \\
SC-LS & LS & 23 & 22/M & 15.0 & 177 & 0.8 & -12.1 & -21.2 & 249.7 & -39.8 & 169 & -51 & 14 & -4.7 & -19.5 & -4.0 \\
O-1A & 25Ori & 7 & 9/16 & 11.2 & 349 & 20.5 & 1.4 & -0.0 & 81.2 & 1.7 & -309 & -119 & -110 & -18.3 & -8.4 & -4.4 \\
O-1B & Gul6 & 7 & 9/A  & 11.2 & 417 & 30.6 & -0.0 & -0.2 & 83.3 & -1.7 & -359 & -169 & -130 & -26.1 & -12.9 & -9.4 \\
O-A & NGC1977 & 7 & 11 & 8.0 & 390 & 31.1 & 1.2 & -0.8 & 83.8 & -4.8 & -324 & -176 & -128 & -25.1 & -16.5 & -8.6 \\
O-A & ONC & 2 & 11 & 2.2 & 403 & 27.2 & 1.4 & 0.8 & 83.7 & -5.5 & -332 & -185 & -134 & -23.4 & -12.9 & -5.9 \\
O-B & $\sigma$Ori & 1 & 18/A & 3.0 & 401 & 31.3 & 1.5 & -0.5 & 84.7 & -2.6 & -341 & -173 & -120 & -26.2 & -15.9 & -7.0 \\
O-MS & LP2439 & 12 & 6/A & 27.0 & 282 & 23.4 & -7.4 & -2.6 & 103.6 & -5.9 & -221 & -176 & -10 & -19.0 & -13.3 & -11.0 \\
O-MS & NGC2232 & 12 & 5/9 & 27.8 & 323 & 25.8 & -4.7 & -1.8 & 96.9 & -4.7 & -264 & -181 & -42 & -20.7 & -13.4 & -10.8 \\
O-P2 & IC348 & 12 & 9/15 & 4.7 & 322 & 15.4 & 4.5 & -6.4 & 56.1 & 32.2 & -289 & 102 & -98 & -16.8 & -6.4 & -7.6  \\
O-P2 & NGC1333 & 12 & 9/15 & 6.0 & 296 & 15.1 & 7.4 & -9.9 & 52.3 & 31.3 & -258 & 102 & -104 & -17.8 & -11.0 & -9.5 \\
 & $\lambda$Ori & 7 & 26 & 5.6 & 400 & 27.5 & 1.1 & -2.1 & 83.8 & 9.8 & -377 & -102 & -83 & -24.7 & -11.5 & -5.6 \\
CH-CU & NGC7058 & 28 & 28 & 55.3 & 362 & -19.3 & 7.3 & 2.6 & 320.4 & 51.2 & -20 & 361 & 6 & -11.0 & -19.9 & -5.9 \\
CH-OR & $\delta$Lyr & 28 & 28 & 32.1 & 352 & -18.4 & 1.2 & -3.1 & 283.6 & 36.9 & 133 & 312 & 93 & -2.9 & -17.0 & -8.5 \\
CH-CI & RSG5 & 28 & 28 & 35.7 & 339 & -7.0 & 3.6 & 1.4 & 303.1 & 45.0 & 52 & 333 & 35 & -6.0 & -5.7 & -4.2 \\
 & IC4665 & 4 & 9/15  & 23.2 & 345 & -13.3 & -0.9 & -8.5 & 266.6 & 5.6 & 284 & 168 & 101 & -3.0 & -17.1 & -8.6 \\
LAC & UBC159 & 25 & 25/14 & 14.5 & 487 & -13.8 & -2.0 & -4.8 & 339.0 & 39.6 & -52 & 465 & -136 & 11.2 & -14.0 & -3.1 \\
 & NGC2451A & 8 & 9/15 & 35.5 & 192 & 23.4 & -21.1 & 15.3 & 115.7 & -38.3 & -58 & -182 & -24 & -27.2 & -14.6 & -12.7  \\
 & NGC3228 & 8 & 9/25 & 30.2 & 484 & 12.9 & -14.9 & -0.7 & 155.4 & -51.8 & 90 & -474 & 38 & -24.7 & -19.5 & -18.5 \\
SER & UPK41 & 21 & 21 & 8.1 & 468 & -9.5 & 2.6 & -8.4 & 279.9 & 0.3 & 397 & 247 & 23 & -0.3 & -16.4 & -14.1 \\
TAU & L1524 & 13 & 13 & 1.6 & 128 & 16.4 & 7.3 & -22.4 & 68.0 & 25.3 & -123 & 15 & -34 & -15.8 & -11.5 & -9.5 \\
T10 & Tr10 & 10 & 9/15 & 56.6 & 432 & 22.0 & -12.5 & 6.5 & 131.9 & -42.6 & -54 & -428 & 4 & -29.1 & -18.9 & -11.0 \\
T10 & Alessi5 & 10 & 9/25 & 70.0 & 395 & 10.6 & -15.4 & 2.5 & 160.8 & -61.1 & 122 & -376 & -14 & -23.1 & -18.5 & -9.6  \\
V-III & CWNU1044 & 18 & 18/27 & 39.8 & 355 & 20.8 & -12.4 & 9.9 & 128.2 & -50.7 & -14 & -352 & -39 & -26.5 & -19.0 & -9.1 \\
V-IV & Cr135 & 8 & 9/25 & 26.3 & 302 & 15.1 & -10.0 & 6.2 & 109.4 & -37.0 & -106 & -277 & -59 & -18.0 & -7.2 & -11.6  \\
V-IV & Cr140 & 8 & 9/25 & 26.9 & 381 & 16.8 & -8.1 & 4.8 & 110.9 & -32.0 & -160 & -342 & -52 & -19.6 & -8.1 & -11.0 \\
V-IV & NGC2451B & 8 & 9/15 & 40.7 & 364 & 15.2 & -9.7 & 4.7 & 116.1 & -38.0 & -110 & -344 & -43 & -18.8 & -8.7 & -12.1 \\
V-IV & NGC2547 & 8 & 9/16 & 32.3 & 387 & 12.8 & -8.6 & 4.3 & 122.5 & -49.2 & -37 & -381 & -58 & -16.1 & -9.8 & -10.9 \\
V-V & Alessi34  & 27 & 27 & 20.6 & 486 & 12.3 & -5.0 & 5.7 & 120.1 & -50.7 & -43 & -475 & -90 & -18.2 & -10.0 & -5.4 \\
V-VI & BH23 & 8 & 9/25 & 28.8 & 438 & 23.1 & -7.1 & 7.4 & 123.4 & -36.3 & -121 & -420 & -8 & -26.3 & -16.5 & -4.0 \\
V-VII & Pozzo1 & 8 & 9/15 & 9.6 & 347 & 17.7 & -6.5 & 9.5 & 122.4 & -47.3 & -43 & -341 & -46 & -20.9 & -15.0 & -2.8 \\
V-O & IC2391 & 4 & 9/25 & 51.3 & 151 & 14.5 & -24.6 & 23.3 & 130.3 & -53.0 & 1 & -150 & -18 & -23.9 & -14.3 & -5.3 \\
V-O & LP2383 & 27 & 27 & 37.9 & 365 & 22.2 & -5.4 & 5.0 & 95.4 & -16.1 & -254 & -247 & -87 & -23.0 & -5.9 & -9.7 \\
CFN & CFN1 & 19 & 19 & 17.8 & 156 & -8.0 & 19.8 & 9.0 & 336.8 & 79.5 & -74 & 129 & 47 & -11.2 & -13.3 & -4.8 \\
\enddata
\tablenotetext{a}{The \citetalias{Kerr23} complex the population is part of, with the sub-region, if relevant: Alessi 20 (AL20), Austral (AUS), Cepheus Flare (CFL), Sco-Cen (SC) broken into the IC 2602 branch (BR), Lower Centaurus-Crux (LC), Upper Sco (US), Upper Centaurus-Lupus (UC), Corona Australis (CR), and Lower Sco (LS); Orion (O) broken into Orion OB1a/25 Ori (1A), Orion OB1b/Belt (1B), Orion B (B), Orion A (A), Monoceros Southwest (MS), and Perseus OB2 (P2), Cep-Her (CH), broken into Cupavo (CU), Orpheus (OR), and Cinyras (CI), Lacerta OB1 (LAC), Serpens (SER), the Taurus Molecular Cloud (TAU), Trumpler 10 (T10), Vela (V) broken into numbered groups from \citet{CantatGaudin19}, and other (O), and Cepheus Far North (CFN)}
\tablenotetext{b}{Age and velocity citations: A (New average of APOGEE RVs), M (MOCAdb), 1 \citep{ZapateroOsorio02}, 2 \citep{Reggiani11}, 3 \citep{ObsHR18}, 4 \citep{Randich18}, 5 \citep{Carrera19}, 6 \citep{Liu19}, 7 \citep{Zari19}, 8 \citep{CantatGaudin20}, 9 \citep{CantatGaudin20b}, 10 \citep{Dias21},  11 \citep{Grossschedl21}, 12 \citep{Kerr21}, 13 \citep{Krolikowski21}, 14 \citep{Poggio21}, 15 \citep{Tarricq21}, 16 \citep{Franciosini22}, 17 \citep{GalindoGuil22}, 18 \citep{He22}, 19 \citep{Kerr22a}, 20 \citep{Kerr22b}, 21 \citep{Ye22}, 22 \citep{Ratzenbock23a}, 23 \citep{Ratzenbock23}, 24 \citep{Wood23}, 25 \citep{GaiaDrimmel23},  26 \citep{Armstrong24}, 27 \citep{Hunt24}, 28 \citep{Kerr24}}
\vspace*{0.1in}
\end{deluxetable*}

For each population, we adopt mean positions and velocities based on bulk measurements from sources listed in Simbad, and check the results against MOCA to ensure consistency. MOCA does not yet consider membership probabilities and flags for complicating factors like binarity, but it nonetheless contains an extensive database of stellar observables, making it useful for verifying that a result reflects the most recent data. Positions and on-sky velocities typically agree across MOCA and other literature sources, but some RVs had no clear consensus. In these cases, we computed RVs using new data from the APOGEE spectrographs \citep{APOGEE} at the Ir\'en\'ee du Pont Telescope \citep{DuPontTelescope} and Sloan \citep{SDSSTelescope} Telescopes, available to us through SDSS-V \citep{Kollmeier26, SDSSDR19}, selecting stars using the SPYGLASS lists around the target cluster. The resulting RVs consistently have $\sigma_{v_r}<0.5$ km s$^{-1}$. For Sco-Cen subgroups, we accepted the MOCA RV values, as these populations have been covered extensively by RV surveys, and their extents are well enough defined by \citet{Ratzenbock23a} that an average across assigned members reliably represents the population. We summarize all literature populations in Table \ref{tab:externalpops}. 

\section{Methods} \label{sec:methods}

Star-forming events vary in their sizes, timescales, and morphologies, ranging from individual molecular clouds to long-lived events with multiple generations connected by stellar feedback, cloud collisions, or other processes. In our dynamic Galaxy, boundaries between star-forming events are not always clear, but related formation can be defined by spatially and temporally continuous formation on scales typical of molecular clouds or filamentary structures, or through formation in material with ages and dynamics consistent with remnant material originating in an earlier generation. Star-forming complexes disperse after formation, often producing several spatially and dynamically distinct subgroups in the present day. Even when these groups do not currently appear related, sufficiently accurate positions, velocities, and ages can be used to trace their locations back in time, revealing common origins. In this section, we outline our traceback methods and techniques for assessing the relatedness of populations with co-spatial or co-moving formation. 

\subsection{Fine Substructure} \label{sec:finess}

The subclustering analysis from \citetalias{Kerr25b} showed that substructure is possible in low-mass associations, but with detection limit of 7 members, fine structure may remain unknown. Massive associations like Sco-Cen have far too much substructure for groups below this detection limit to draw attention, but in these low-mass associations, subtle substructure may be both detectable and important for understanding their star formation histories. If such substructure is present, it may indicate that stars originated in star-forming events with distinct ages and dynamics, which may produce unrealistic bulk traceback if not separated.


For an ultra-low mass population to be distinct enough to report, we require ages and transverse velocities that differ from other components in the association, satisfying either $\Delta v_T > 2$ km s$^{-1}$ or $\Delta \tau > 5$ Myr. A population at these limits would form $\sim$10 pc from its parent population in its rest frame, signifying an offset near the present-day average half-mass radius for the low-mass populations in \citetalias{Kerr25b}. We define the populations by hand as described in Appendix \ref{app:microclusts}, using quiver plots to identify compact overdensities with closely aligned velocity vectors. Upon identifying a potentially distinct cluster, we produce a color-magnitude diagram (CMD) and average velocity to assess whether the population is distinct by our metrics. 

Four populations emerge as \added{potential} hosts of distinct substructure: ScuN, AndS, OphSE, and TOR1B. All identified groups meet our 5 Myr age difference metric, and one also meets our velocity metric. We define each population and discuss their properties in Appendix \ref{app:microclusts}. \added{Using false positive tests, we find that the substructure in ScuN and OphSE is robust, while the features in TOR1B and AndS are more tentative and require follow-up observations for confirmation.} Two groups have masses of only 1.6 M$_{\odot}$, \added{including one robust structure}, and we discuss the implications of this in Section \ref{sec:disc_finess}.






\subsection{Traceback}

We trace the positions of stellar populations back in time with the \texttt{galpy} module \citep{Bovy15}, using the \texttt{MWPotential2014} galactic potential and \citet{Hogg05} solar motion. For literature populations, we use the bulk motions listed in Table \ref{tab:externalpops}. For the low-mass associations, we perform traceback for all members with non-\textit{Gaia} RVs with $\sigma_{RV}<1$ km s$^{-1}$ and no evidence for unresolved binarity from Gaia ($RUWE>1.2$; \citealt{Bryson20}) or resolved binarity in \citetalias{Kerr25b}, and average their positions at each time step. We only consider the fine substructure in ScuN, as the velocities of new  subgroups in AndS and TOR1B do not differ substantially from the bulk velocities, and OphSE-B lacks high-quality RVs (see Appendix \ref{app:microclusts}). We trace the low-mass and literature populations back 60 Myr, which is the maximum plausible age for any low-mass association in \citetalias{Kerr25b}. 

The \texttt{galpy} module provides results in galactocentric cartesian coordinates, which we convert into a co-rotating frame similar to the curvilinear heliocentric coordinates ($\xi^\prime$, $\eta^\prime$, $\zeta^\prime$) defined in \citet{Asiain99}. However, rather than accounting for galactic rotation using a standard solar orbital frequency \citep{MiretRoig20}, we compute ($\xi$, $\eta$, $\zeta$) by converting from galactic (X, Y, Z) to cylindrical coordinates, subtracting the average $\phi$ and $\rho$ for the populations of interest, and converting back to heliocentric cartesian coordinates with the Galactic Center locked to the right. This frame minimizes motion relative to the average for visualization purposes. 

\subsection{Identifying Connected Star-Forming Events} \label{sec:connsfe}

Populations that originate in the same molecular cloud converge on a common origin during dynamical traceback, making them easy to relate to one another. However, a common formation origin may not be expected for populations that form through triggering by a collision with another gas structure, or as part of an expanding bubble  \citep[e.g.,][]{Tan00, Elmegreen11}. For bubble-driven expansion, stars may form with outbound velocities and younger ages compared to the parent complex, and will normally have a close pass with other associations at some point during their traceback. Cloud-cloud collisions, which may be a significant driver of star formation in galactic disks, typically produce two dominant velocity components consistent with each colliding cloud in the absence of gravitational binding, with formation starting near the time of the collision \citep{Wu17, FortuneBashee24}. In both cases, velocities may differ from other components of the complex by several km s$^{-1}$ \citep[e.g.,][]{Kerr25}. It is therefore not straightforward to assess relationships between associations in an automated way, so we define a series of metrics to flag associations as potentially connected, meriting further inspection. 

We first compute the separation in both space and velocity coordinates from each low-mass association to both literature associations and other low-mass associations. We compute these values for times within 20\% of the best fit PARSEC ages, accounting for the systematic uncertainties that dominate most literature ages \citep[e.g.,][]{Herczeg15, Wood23}. Any population within 200 pc with velocities within 5 km s$^{-1}$ during that time interval is marked as co-moving at formation, while populations within 100 pc are marked as co-spatial. The co-moving category groups populations that may have formed as part of a larger comoving and coherent gas structure, even if their formation was spatially disconnected. The 5 km s$^{-1}$ limit aligns with typical velocity spreads in large star-forming complexes like Sco-Cen \citep{Kerr21}, as well as the maximum velocities seen feedback-driven and star-forming expanding shells like in $\lambda$ Ori or the Circinus Complex \citep{Armstrong24, Kerr25}. The co-spatial category identifies populations that may have been related through higher-velocity triggering scenarios, such as through particularly fast expanding bubbles or cloud-cloud collisions, which can have velocities of 10-15 km s$^{-1}$ \citep{Tasker09, Li18}. Both conditions are generous, so additional analysis is necessary to assess the plausibility of any potential common origin. 

\begin{deluxetable*}{c|c|c|c|c}
\tablecolumns{5}
\tablewidth{0pt}
\tabletypesize{\scriptsize}
\tablecaption{Literature populations that we find to be co-spatial or co-moving with our low-mass young associations at formation. We group low-mass associations and their matches by overlap in their co-spatial and co-moving lists, roughly defining ``families'' discussed later in this paper. Populations in red lack ages and velocities consistent with formation in a common cloud or formation influenced by a bubble model from Sec. \ref{sec:clfams_coher}, so their connection to the parent family may be spurious.}
\label{tab:commongroups}
\tablehead{
\colhead{Family} &
\colhead{Low-Mass} &
\multicolumn{3}{c}{Literature Population Matches} \\
\colhead{} &
\colhead{Assoc} &
\colhead{only co-moving} &
\colhead{only co-spatial} &
\colhead{both}}
\startdata 
Local &  CMaN & $\chi^1$For, THA, Platais8, LS  & $\delta$Sco, eLup, CrA, \textcolor{red}{NGC1333}, \textcolor{red}{RSG5}, CFN1 &  IC2602, $\sigma$Cen\\
 &  Theia 78-1 & $\chi^1$For, Platais8 & THA, IC2602, $\sigma$Cen, $\delta$ Sco, e Lup, CrA, LS, \textcolor{red}{RSG5}, CFN1 &  \\
  &  Theia 78-2 & Platais8 & THA, IC2602, $\sigma$Cen, $\delta$Sco, eLup, CrA, LS, \textcolor{red}{NGC1333}, \textcolor{red}{RSG5}, L1524, CFN1 & \\
& AqE & &  $\chi^1$For, Platais8, \textcolor{red}{RSG5}, CFN1 & CrA \\
&  ScuN-AE &   & $\chi^1$For, THA, Platais 8, \textcolor{red}{RSG5}, CFN1 & CrA\\
&  ScuN-Cen &  &  & CrA \\
&  ScuN-Halo &  Lower Sco, $\delta$ Sco & & CrA \\ \hline
Orion &  AriS & 25Ori, \textcolor{red}{IC348} & Gul6, NGC1977, ONC, $\sigma$Ori, $\lambda$Ori, \textcolor{red}{L1524} &  \\
&  TOR1B & 25Ori,\textcolor{red}{IC348}, \textcolor{red}{NGC1333} & NGC1977,$\sigma$Ori,$\lambda$Ori &  \\
&  VulE &  & Gul6, NGC1977, ONC, $\sigma$Ori, LP2439, NGC2232, \textcolor{red}{NGC1333}, $\lambda$Ori, \textcolor{red}{L1524} &  \\ \hline
Vela  &  AndS & \textcolor{red}{IC348} & Cr135, Cr140, NGC2451B, NGC2547, LP2383 &  \\ 
 &  Theia 78 & LP2383 &  &  \\ \hline
Other &  SCYA-79-0 &  &  & RSG5 \\
  &  SCYA-79-1 &  & THA, IC2602, $\sigma$Cen, eLup, CrA, CFN1 & RSG5 \\
  &  SCYA-54-0 & NGC7058 & Platais8 & $\alpha$Per \\
  &  SCYA-54-1 & NGC7058 &  & $\alpha$Per \\
  &  CasE & Al20,ASCC127,CFN1 &  &  \\
  &  LeoC &  & NGC1977, $\sigma$Ori, LP2439 &  \\
\enddata 
\end{deluxetable*}

\section{Results} \label{sec:results}

\subsection{Connections Between Low-Mass Associations}

Our traceback reveals three pairs of low-mass populations that were both co-spatial and co-moving during the formation of at least one of the two components: CMaN and Theia 72, AqE and ScuN, and LeoE and LeoC. Here we assess whether their ages and positions at formation are consistent with an origin in a common star-forming cloud. 

\begin{figure*}
\centering
    \includegraphics[width=18cm]{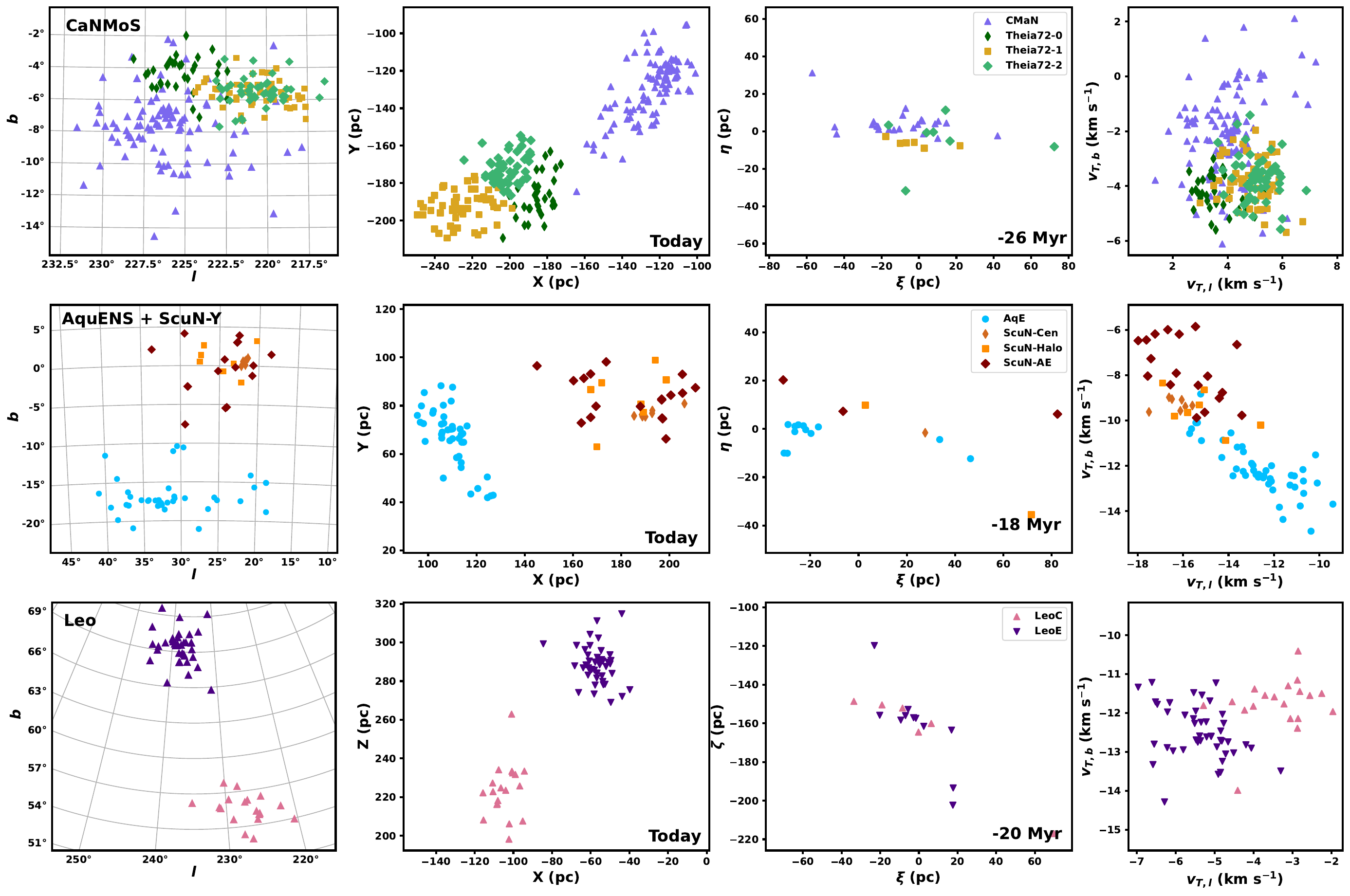}
\caption{Pairs of low-mass stellar populations that form larger complexes: CaNMoS (CMaN + Theia 72; top), AquENS+ScuN-Y (AqE+ScuN; middle), and Leo (LeoE + LeoC; bottom). In the left column, we show their on-sky distributions in $l$ vs $b$ galactic coordinates. We then show their present-day distribution in X vs Y galactic coordinates in the second column, or X vs Z coordinates for Leo, which separates more prominently in Z. In the third column, we show their distributions in galactic cartesian coordinates near the time of their formation after traceback, which contains fewer points due to the need for high-quality RVs. In CaNMoS and Leo, the distributions of the component populations at formation overlap. Across AqE and ScuN, the convergence is less clear, as only AquENS (AqE + ScuN-AE in ScuN) nearly overlap at formation. ScuN-Cen  and ScuN-Halo are dynamically similar, but do not form a common star formation event with AqE. In the final column, we show the distributions of all three association pairs in transverse velocity, where velocity distributions for subcomponents consistently overlap.}
\label{fig:combpops}
\end{figure*}

\subsubsection{Canis Major North and Theia 72}

CMaN and Theia 72 both have bulk ages of $\sim$26 Myr \citep{Kerr25b}, and our traceback places their centers within only 18 pc of each other at formation, with velocities within 4 km s$^{-1}$. We show the distribution of CMaN and Theia 72 members in spatial coordinates in the present day and at formation in the left two panels of the Figure \ref{fig:combpops}'s top row, showing that while the populations span nearly 200 pc in the present day, all components likely overlapped at formation. Their present-day separation appears to be driven by a highly asymmetric velocity dispersion along a largely linear axis pointed away from the sun, which we discuss in Appendix \ref{app:localvels}. We show their velocities and on-sky spatial distributions in the right two columns of that top row, revealing clear overlap. The co-spatial formation, similar velocities, and co-eval formation of Theia 72 and CMaN make it unlikely that their formation is unrelated. We therefore classify both populations as subgroups within a common ``CaNMoS'' Association (Canis Major North - Monoceros South).

This combination of co-spatial formation and a 4 km s$^{-1}$ velocity difference between primary components is similar to the Taurus Molecular Cloud, where \citet{Krolikowski21} reveals that stellar velocities associated with clouds like L1495, L1524, and L1527 differ from those associated with L1536 and B213 by about the same magnitude on average. Taurus is also similar in size to CaNMoS, with 532 adopted members from \citet{Luhman22} compared CaNMoS's estimated membership of 432 stars. Taurus may therefore represent a young analog to CaNMoS. 

\subsubsection{Aquila East and Scutum North}

Aquila East and Scutum North are also both co-spatial and co-moving at formation, with bulk position and velocity separations of $\sim3$ km s$^{-1}$ and $\sim$50 pc at the times of their bulk ages. They are therefore more dynamically similar than the components of CaNMoS, but without as clear of a co-spatial origin. The oldest of the three ScuN components described in Section \ref{app:scunss}, ScuN-AE, consists of stars with velocities and ages consistent with dispersal from AqE after its formation. As shown in the middle row of Figure \ref{fig:combpops}, our traced-back stellar samples for ScuN-AE and AqE nearly overlap at formation, with an median separation of 19 pc. This suggests that they comprise a single star-forming event, which we refer to as the ``AquENS'' Association (Aquila East - Northern Scutum) for ease of reference. 

The other ScuN components, ScuN-Cen and ScuN-Halo, have weaker connections to the two earlier generations, as traceback reveals median separations to AqE of 70 and 95 pc, respectively. Their ages are also inconsistent with AqE, with the older ScuN-Halo differing from AqE by $\sim$6 Myr. These groups have limited RV coverage, and those we have produce a spatial scatter of nearly 100 pc at formation, so deeper coverage may be necessary to better characterize their motions. However, none of our existing measurements support co-spatial formation with the proposed AquENS node, largely ruling out formation in a common cloud. However, ScuN-Cen and ScuN-Halo remain dynamically similar to AquENS, with velocities separated by 2.1 and 3.2 km s$^{-1}$, respectively, so it is possible that the gas they formed from shares some common origin. 

ScuN-Cen and ScuN-Halo have a closer relationship to each other than to AquENS, with a separation of 20 pc and 2.5 km s$^{-1}$ when ScuN-Cen forms. Their velocities are convergent, which does not support expansion from a common structure, but with only 4 RV measurements between them, this is uncertain. Their relationship to each other is therefore inconclusive, but we refer to them as Scutum North-Young (ScuN-Y) for future reference. 

\subsubsection{The Leo Association} \label{sec:Leonode}

Both Leo populations have unusual near-vertical velocities relative to the Galactic plane, with $W$ velocity components at the time of LeoC's formation of 25.5 km s$^{-1}$ in LeoE and 21.5 km s$^{-1}$ in LeoC. Young associations with velocities that differ from solar velocity by more than 20 km s$^{-1}$ are not unusual in the solar neighborhood, but \citetalias{Kerr23} shows no associations out of the 116 studied with an average $v_{T,b}$ exceeding the vertical velocity of either Leo population. The $b$ velocity component is vertical to the Galactic midplane for stars near $b = 0$ where most young associations are located. The lack of associations with a $v_{T,b}$ equaling or exceeding LeoE or LeoC's vertical velocities therefore indicates that those high velocities are at least rare, if not unique within the solar neighborhood. 

The traced-back separation between LeoE and LeoC reaches a minimum of only 9 pc $\sim20$ Myr ago, which is within uncertainties of LeoE's $21.7\pm2.1$ Myr age. They therefore appear to share a common origin within a typical cloud scale of each other, while also sharing velocities consistent with divergence after LeoE's formation. This is consistent with an origin in a common structure, which we call the Leo Association, or Leo for short.

\subsection{Connections Between Low-Mass Associations and Literature Populations} \label{sec:conn-lit}

\begin{figure*}
\centering
    \includegraphics[width=17.5cm]{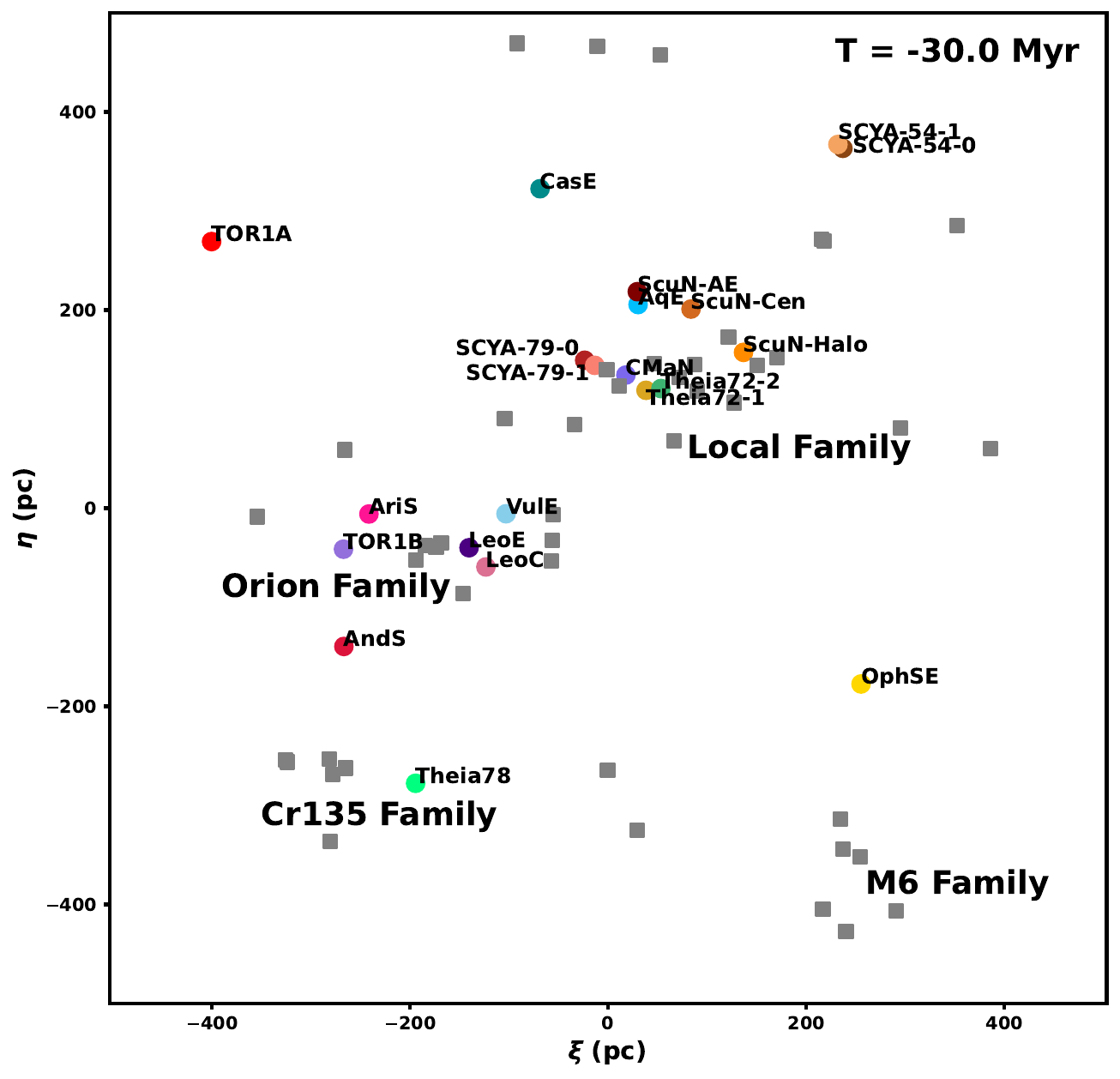}
\caption{Traceback of all associations in both the literature sample (grey squares) and low-mass sample (colored circles). Here we present one timestep 30 Myr before present, where overdensities of populations connected to the Local Bubble, Orion-Eridanus Superbubble, and the Cr135 and M6 families from \citet{Swiggum24} are evident. The online-only version shows the trajectories of these populations and their progenitor clouds over 60 Myr. This version of the figure can be interacted with in 3 dimensions by zooming, panning, and rotating. Mousing over a population provides basic information like position and age. We also include a button to limit the figure to only groups that have formed at a given time.}
\label{fig:bigtraceback}
\end{figure*}

In Table \ref{tab:commongroups}, we summarize the literature populations that are co-moving or co-spatial with each low-mass association, revealing that many have overlapping lists. AqE, ScuN, and CaNMoS all share position and/or velocity space with populations in the Sco-Cen Association and Austral Complex, in addition to populations like L1524 in Taurus and CFN-1. AriS, TOR1B, and VulE all relate to populations in \citetalias{Kerr23}'s Orion-Perseus Complex, which includes Orion OB1 and Monoceros Southwest (see \citetalias{Kerr21}). Finally, AndS and Theia 78 relate to components of the Vela complex, especially those grouped into Vela group IV in \citet{CantatGaudin19}. We show the traceback of all populations in Figure \ref{fig:bigtraceback}, displaying the -30 Myr time step in text and the full time series online-only. Many young associations reside in clear overdensities at the -30 Myr time step, similar to the families proposed by \citet{Swiggum24}. 

We find that most low-mass populations share a plausible connection to a larger complex. CaNMoS (CMaN and Theia 72), ScuN-Y, TOR1B, and AriS all have velocities, positions, and ages consistent with material swept up in parent bubble structures, and SCYA-79 is consistent with membership in the Cinyras association. Theia 78 and AquENS lack velocities consistent with a bubble, but form where and when bubbles are likely to influence star formation, such as through feedback-driven triggering. Four additional populations show some evidence of interactions with complexes, namely VulE and the Leo Association (LeoC and LeoE), which may have formed while interacting with Orion's progenitor clouds, and OphSE, which has outbound velocities from the M6 Family \citep{Swiggum24}. AndS, TOR1A, CasE, and SCYA-54 are the only populations with no clear connections to literature populations, and only CasE and SCYA-54 form within a volume where our literature sample is complete enough to assess their formation environment.

\begin{figure*}
\centering
    \includegraphics[width=17.5cm]{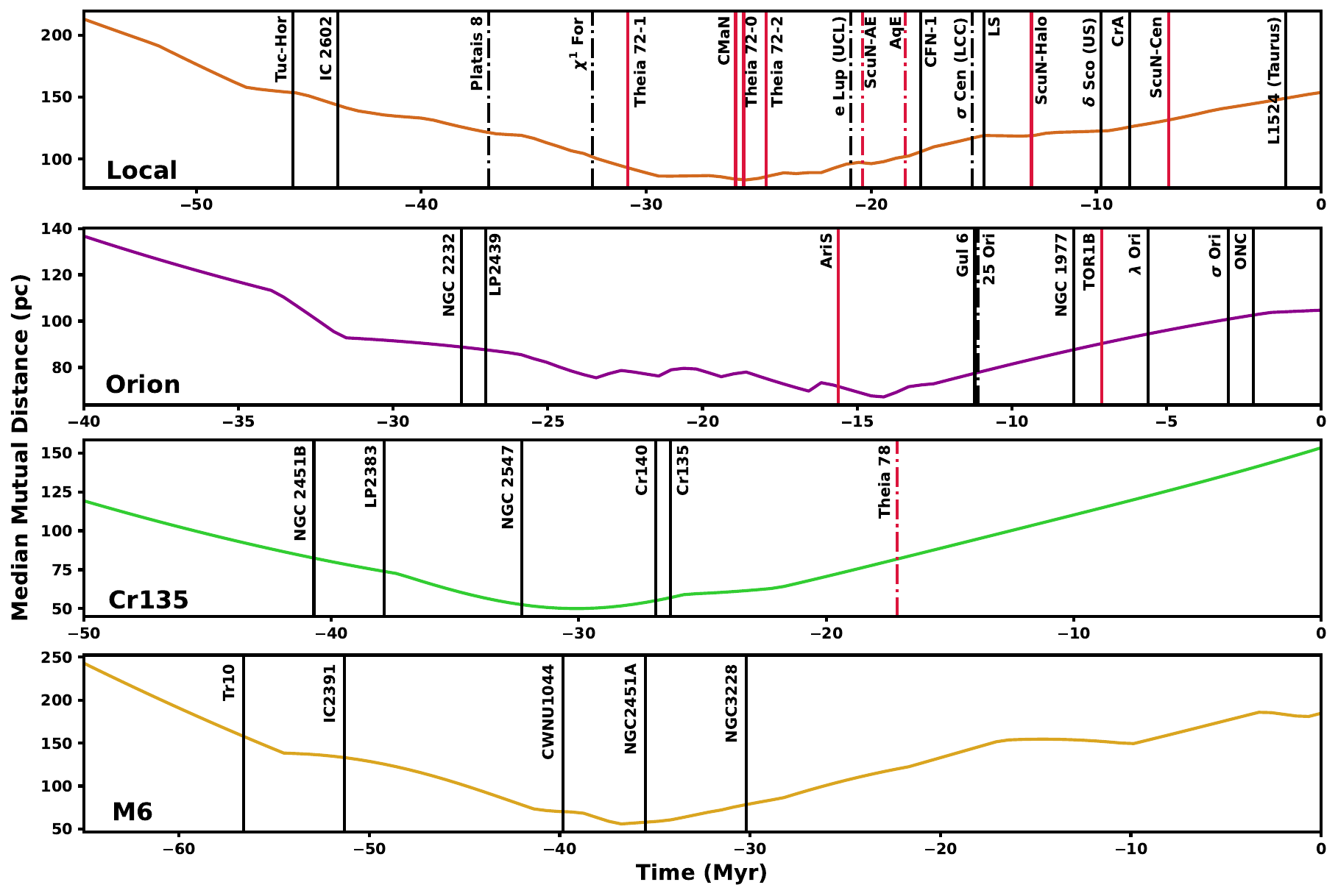}
\caption{Formation and traceback histories for the Local and Orion Families, which we associate with the Local and Orion-Eridanus Bubbles, in addition to the Cr135 and M6 Families from \citet{Swiggum24}. The colored curves show the median mutual distance between members of the family, with the minimum indicating a potential start time for expansion. The times of formation for literature populations (black) and low-mass associations (red) in each family are marked with vertical lines. Solid lines indicate populations we relate to the early expansion of a feedback-driven bubble or star formation in the bubble itself, indicated by outbound velocities and a position consistent with a presumed bubble edge. Populations consistent with triggering through a collision with a bubble but not bubble membership are marked with broken vertical lines, indicated by a formation site near a presumed bubble edge, but without outward-oriented velocities.}
\label{fig:familyhistory}
\end{figure*}

We group populations associated with AqE, ScuN, and CaNMoS into the Local Family, a group of young associations with a potential relationship to the Local Bubble, and assign AriS and TOR1B to the newly-defined Orion Family, related to the Orion-Eridanus Superbubble. The Cr135 and M6 Families from \citet{Swiggum24} also appear in our literature sample, but they have weaker connections to our low-mass set. To confirm that populations in the same generously-defined co-spatial and co-moving set are part of the same family, in this section (and Appendix \ref{app:conn_supldisc}) we assess whether the locations and velocities of populations at formation are consistent with a common star-forming cloud or feedback-driven bubble produced by a prior generation. We mark populations with rejected membership in a potential family in Table \ref{tab:commongroups}, and summarize the star formation history of populations with plausible connections to the Local, Orion, Cr135, or M6 Families in Figure \ref{fig:familyhistory}. In all four families, the first stellar generations form $\sim10-20$ Myr before the minimum in median mutual distance between members. This offset may be caused by the time between star formation and the first supernovae, or formation in accelerating clouds, resulting in present-day dispersal velocities that overestimate the average velocity.

\subsubsection{CaNMoS, AquENS, and ScuN-Y} \label{sec:local} 

\begin{figure*}
\centering
    \includegraphics[width=16cm]{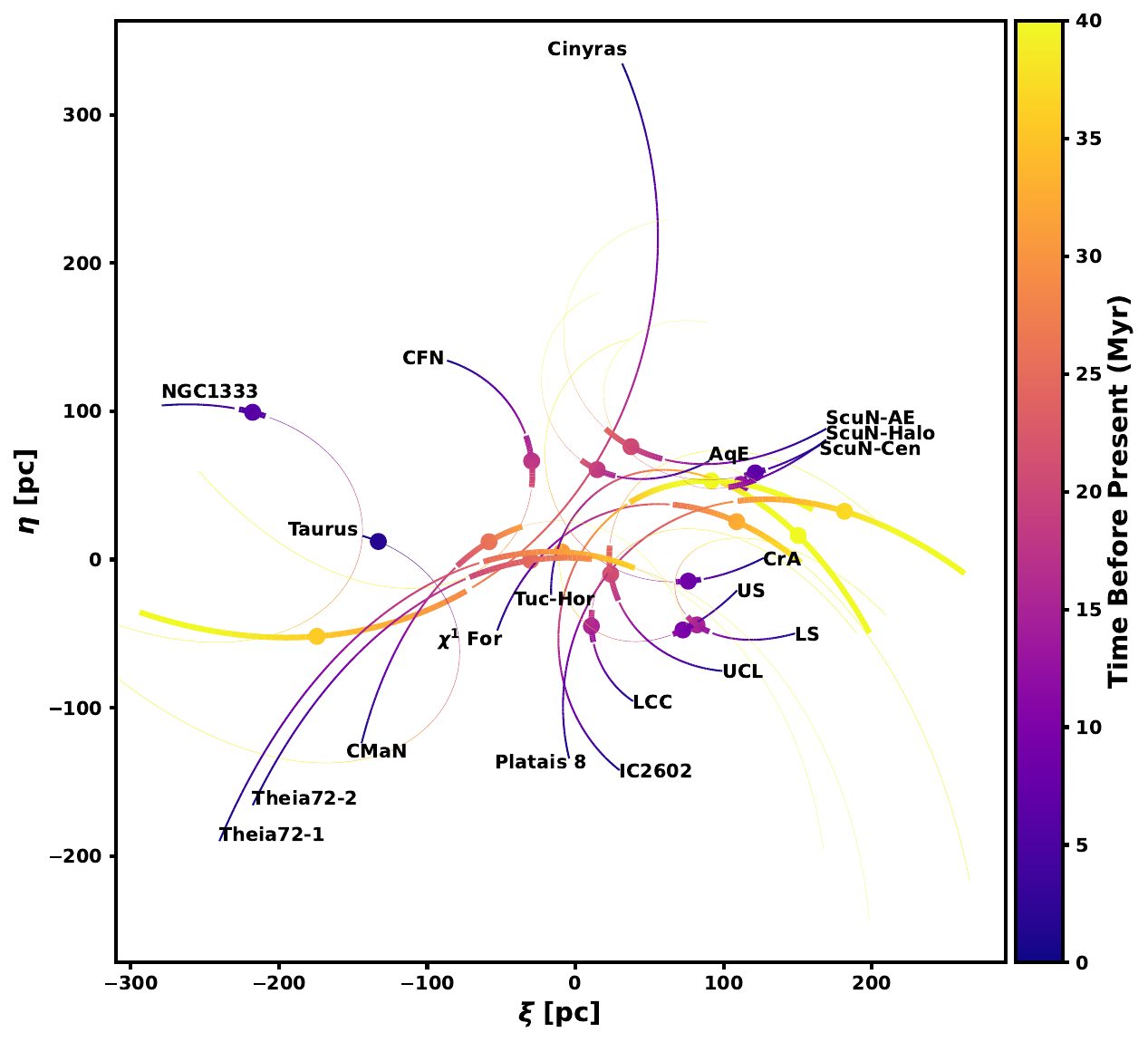}
\caption{Dynamical traceback of CMaN, Theia 72, AqE, ScuN and literature associations that were co-moving or co-spatial with any of them at formation, showing divergence from a close configuration between 20 and 30 Myr ago to their widely distributed positions in the present day. Curves show the trajectories of populations colored by time, with dark colors for times closer to the present day, shown in $\xi$ vs $\eta$ co-rotating galactic coordinates (XY coordinates with the Galactic center locked to the right and centered on the local average). Dots indicate formation sites based on the best fit age, with the thick bar around them indicating a 20\% systematic age uncertainty. The lines are thin for times before this interval and thick after it, indicating the added uncertainty inferring gas motions using stars. Typical velocity uncertainties are $<0.5$ km s$^{-1}$ \citep{Kerr25b}, corresponding to a positional uncertainty of $\sim$0.5 pc/Myr before present. The divergence is centered on the Local Bubble, and components of CaNMoS (CMaN and Theia 72) and ScuN-Y (ScuN-Cen and Halo) form with largely outward velocities, consistent with formation on its edge, while AqE and ScuN-AE formed with inbound velocities, potentially through a collision between their progenitor cloud and that bubble.}
\label{fig:canmostraceback}
\end{figure*}

In Figure \ref{fig:canmostraceback}, we present a simplified visualization of the traceback results in Figure \ref{fig:bigtraceback} covering all populations that are co-moving or co-spatial with at least one component of the CaNMoS node at formation in our traceback (including all populations associated with AqE and ScuN). All populations show divergence in the present day, with most sharing a roughly 100 pc-wide volume between 20 and 30 Myr ago. NGC 1333 in Perseus and RSG 5 are notable outliers due to their unusual trajectories and presence in the co-spatial and co-moving lists for other low-mass associations. When CaNMoS forms, NGC 1333 is co-spatial with it in our traceback, so some continuity is possible \citep[e.g.][]{Koda23}, but NGC 1333 forms long after, leaving time for interactions with other gas structures and bubbles. Pre-formation traceback should therefore be treated with caution. This also applies to L1524 in Taurus, but its closest approach to the rest of the complex is much more recent. RSG 5 in Cinyras is distinct in its velocities, which differ from CMaN by 18.5 km s$^{-1}$. It was within only 38 pc of CaNMoS at formation, but formed 10 Myr before it \citep{Kerr24}. RSG 5 is not the oldest component of Cinyras, so even if the age of Cinyras is being underestimated, it likely started forming more than 200 pc from other coeval populations in this set, making a direct connection improbable. We therefore exclude RSG 5 and NGC 1333 from this analysis. 

All remaining populations have motions consistent with a common origin, including Taurus, Sco-Cen, IC 2602, and components of the Austral Complex. The connectedness of these populations is not a new observation, as \citet{Zucker22b} proposed that the expansion of the Local Bubble, initiated by stellar feedback from the UCL and LCC groups in Sco-Cen, powers star formation in later Sco-Cen generations and the Taurus Molecular Cloud. \citet{Swiggum24} made a similar observation, grouping CaNMoS and its co-spatial and co-moving relatives with $\alpha$ Persei, one of three cluster ``families'': groups of stellar populations that are currently widely distributed, but share a close configuration in the past that suggests a common origin. Our list of CaNMoS relatives excludes the namesake $\alpha$ Per Cluster, but this cluster is a visible outlier within its family. The minimum size of the $\alpha$ Per Family (calculated as the median distance between each cluster and the geometric center) was 88 pc in \citet{Swiggum24}, far larger than our set of CaNMoS relatives that excludes it, with a minimum size of 60 pc. At $\sim80$ Myr old \citep{GalindoGuil22}, the $\alpha$ Per cluster is also much older than even Tuc-Hor, the oldest population that we associate with CaNMoS at 45.7 Myr \citep{Wood21}. Like \citet{Zucker22b}, \citet{Swiggum24} connects the $\alpha$ Per family to the Local Bubble. Our traceback similarly shows a clear divergence signature centered on the Local Bubble that may be related to its expansion, but our traceback does not support $\alpha$ Per's inclusion. We hereafter refer to the component of the $\alpha$ Persei Family associated with CaNMoS and the Local Bubble as the ``Local Family'', following the terminology in \citet{Swiggum24}. 

The low-mass populations of CMaN, Theia 72, ScuN, and AqE all have positions and velocities at formation that support a relationship with the literature populations that comprise the Local Family. Most of CaNMoS forms 7 Myr after the co-moving populations of $\chi^1$ For and Platais 8 with a velocity directed away from both, indicating dynamical relatedness to the generation that directly preceded its formation. ScuN-Y also forms with young ages and radially-oriented velocities relative to the rest of the family. These patterns are typical indicators of triggering or formation in a feedback-driven shell. The relationship between AquENS and the Local Family is less clear, as they have slightly convergent velocities inconsistent with bubble gas. However, the proximity of AquENS to populations in the family at formation and relatively young age indicate that a feedback-driven bubble may have influenced its formation. We explore whether the ages and dynamics of these associations can be explained through a multi-generational sequence of bubble-influenced star formation in Section \ref{sec:bubble-local}, and discuss the potential origins of their internal velocity spreads in Appendix \ref{app:localvels}.



\subsubsection{SCYA-79 and Cinyras}

\begin{figure*}
\centering
    \includegraphics[width=10.2cm]{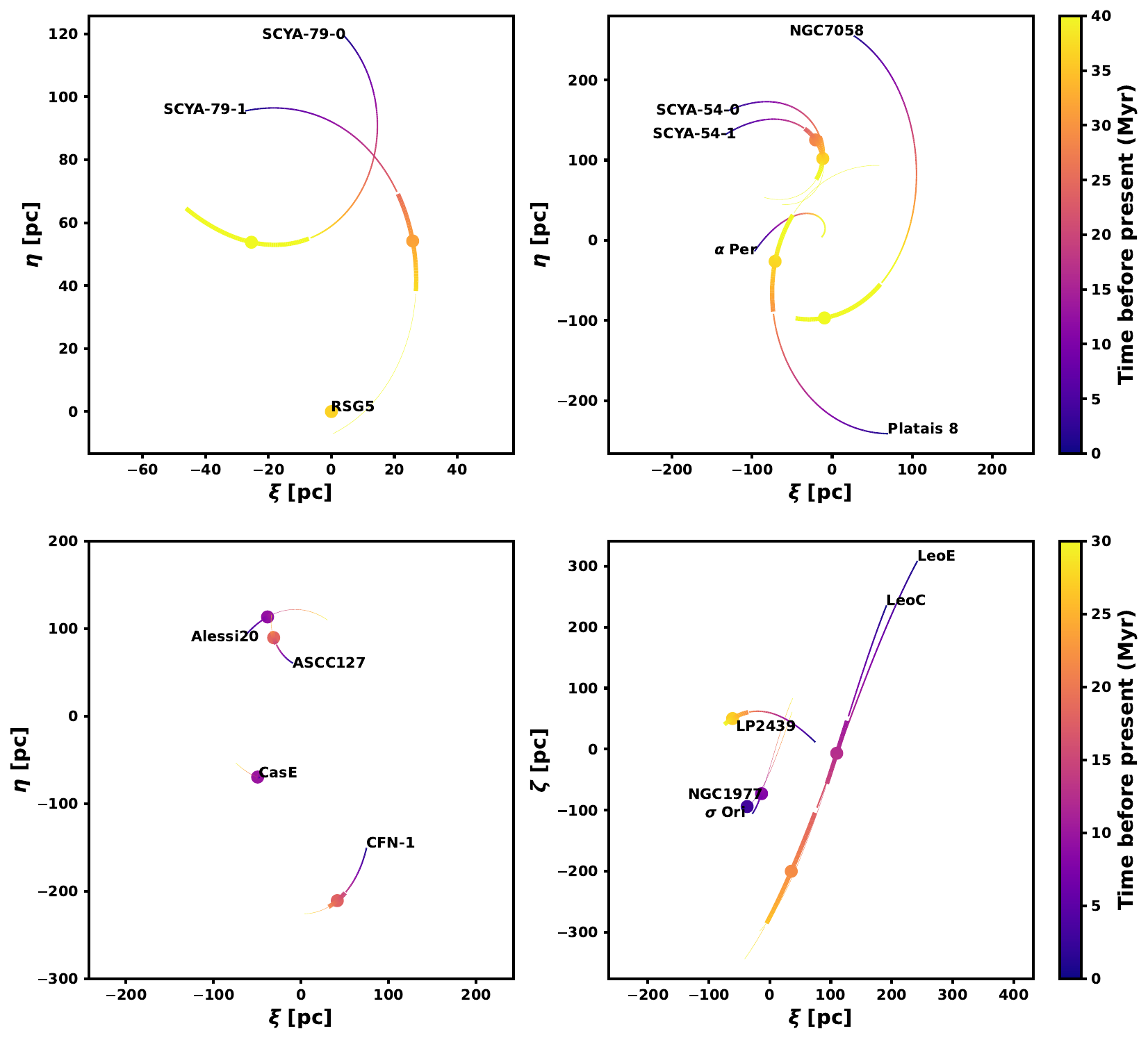}
    \includegraphics[width=7.7cm]{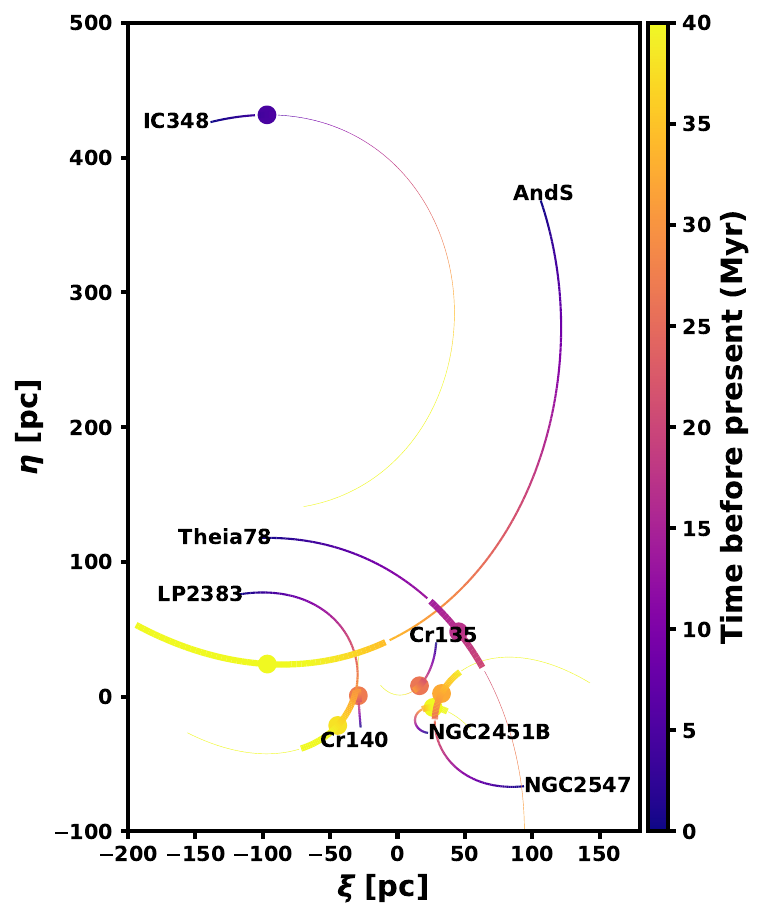}
\caption{Same as Fig. \ref{fig:canmostraceback}, but for the populations in our low-mass set outside the Local Family and Orion. In the top row at left, we show SCYA-79 (left) and SCYA-54 (right) relative to their co-moving and co-spatial sets, traced back 60 Myr and sharing the color bar on the right. In the bottom row at left, we show CasE (left) and the Leo Association (right), traced back 30 Myr and sharing the color bar on the right. At right, we show stellar populations co-spatial or co-moving with AndS or Theia 78 at formation, corresponding roughly to the Collinder 135 family from \citet{Swiggum24}. Theia 78 forms outbound from that family, but with a notably non-radial velocity, giving it an uncertain connection, while AndS has a velocity mismatch with the rest of the family that makes relatedness unlikely. SCYA-79 shows velocities and ages consistent with an origin in the Cinyras Association, and it is the only population in this set with a plausible connection to a literature population. All others show uncertain connections with other association, although LeoE and LeoC have motions consistent with formation while entering a portion of the galactic plane with known clouds.} 
\label{fig:othertraces}
\end{figure*}

Assuming the isochronal ages from \citetalias{Kerr25b}, SCYA-79 was both co-spatial and co-moving with RSG 5 in Cinyras at formation, within 43 pc for SCYA-79-1, and within 59 pc for SCYA-79-0. We show the motions of the SCYA-79 subgroups relative to RSG 5 in the top-left panel of Figure \ref{fig:othertraces}, which shows divergence between both components and RSG 5 in the present day, and indicates that SCYA-79-1 forms outbound from a close approach within 14 pc about 50 Myr ago. SCYA-79-0 has a slightly inbound velocity at formation, however it has a particularly high 0.9 km s$^{-1}$ velocity dispersion across 4 useful members. This propagates to a 45 pc position error after 50 Myr, which would nearly erase its separation from RSG 5. The velocity differences between RSG 5 and these subgroups are 2.5 and 4.6 km s$^{-1}$, respectively, within the approximately 4 km s$^{-1}$ velocity span in known Cinyras subgroups. 

The ages of SCYA-79 are also consistent with those in Cinyras, but the weak separation between SCYA-79's PMS and field sequence produced uncertain isochronal ages in \citetalias{Kerr25b}. The lithium depletion age of $35.5^{+12.6}_{-17.0}$ Myr for the entire population is more reliable, since it does not require main sequence stars with uncertain membership, but it remains uncertain. All age solutions from \citetalias{Kerr25b} are nonetheless consistent with the 29-43 Myr range for the Cinyras complex from \citetalias{Kerr24}, as well as the 35.7 $\pm$ 1.5 Myr PARSEC age for RSG 5. This, combined with the divergent velocities relative to RSG 5 after formation makes SCYA-79 a plausible outlying component of the Cinyras complex. An analysis that considers the full range of subgroups in Cinyras and improved velocity coverage in SCYA-79 could strengthen that connection.

\subsubsection{Orion-Eridanus} \label{sec:orieri}

\begin{figure*}
\centering
    \includegraphics[height=11cm]{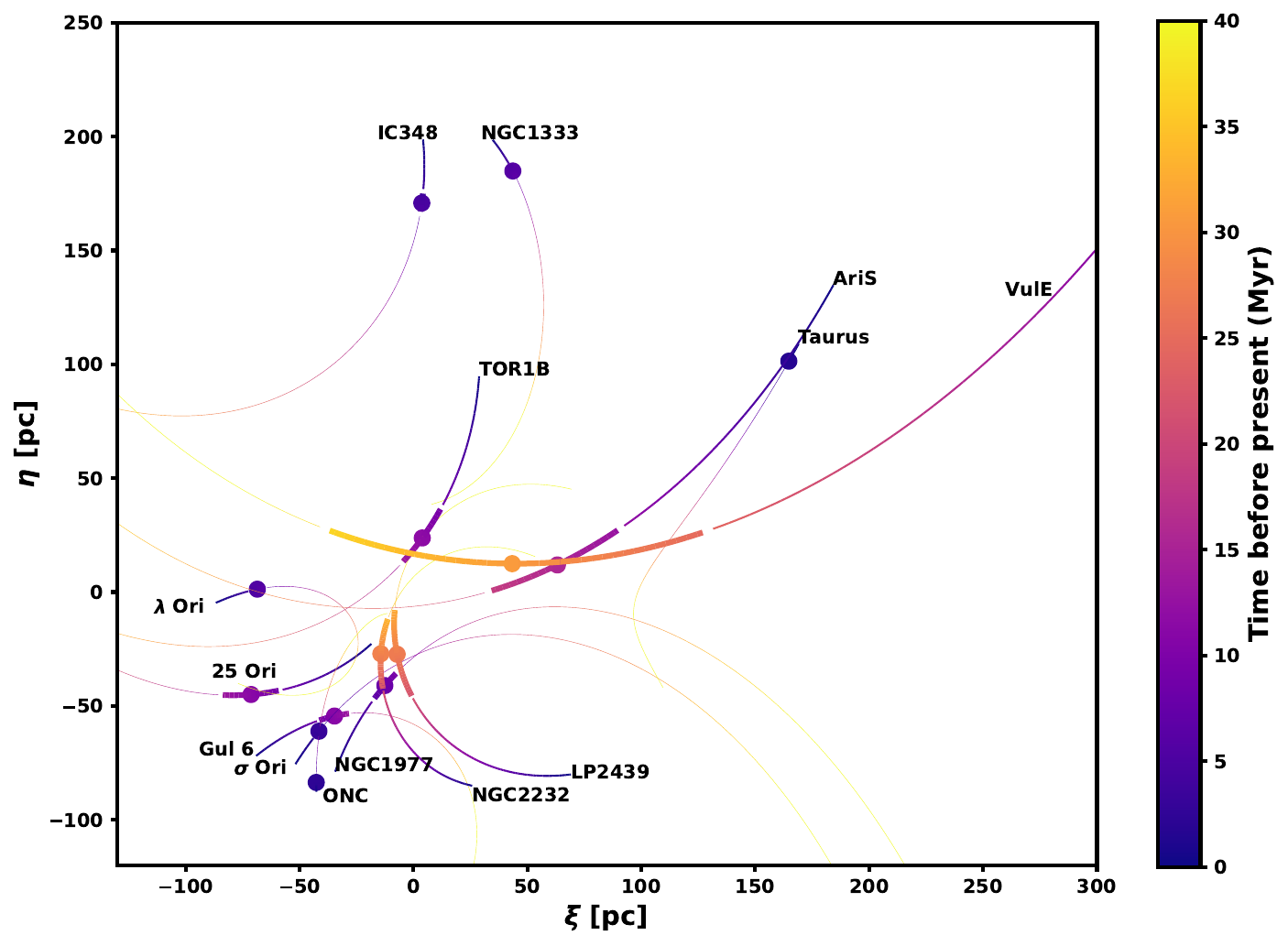}
\caption{Same as Fig. \ref{fig:canmostraceback}, but for the populations co-spatial or co-moving with AriS, VulE, or TOR1B at formation, which comprise the Orion Family. Most populations exhibit divergence, and AriS and TOR1B form outbound from the center of Orion, consistent with an origin in the enclosing Orion-Eridanus Superbubble. VulE's connection to Orion is less clear, but it may have interacted with the progenitor clouds to NGC 2232 and LP 2439. }
\label{fig:oriontraceback}
\end{figure*}

Most co-moving and co-spatial populations to AriS, TOR1B, and VulE at formation are associated with the Orion Complex: ASCC 16 (25 Ori), Gulliver 6 (in Orion OB1b), the ONC and NGC 1977 (Orion A), and $\lambda$ Ori, in addition to NGC 1333 and IC 348 in Perseus and NGC 2232 and LP2439 in Monoceros Southwest (MSW), which \citetalias{Kerr23} includes in the Perseus-Orion Complex. L1524 in Taurus, which we include in the Local Family, is co-spatial with VulE and AriS when they form, but like for NGC 1333's inclusion in the Local Family, this connection assumes that current stellar motions reflect those of the progenitor cloud 15 Myr ago. We therefore exclude it from our analysis. We show the motions of all populations in this set in Figure \ref{fig:oriontraceback}, revealing a similar dispersal pattern to the Local Family. 

The Orion Nebula Complex has been tied to the Orion-Eridanus Superbubble, a large, multi-shelled structure spanning up to 250 pc \citep{Bally08-ori, Joubaud19}. This structure has been attributed to feedback originating in Orion OB1, which contains most of the famous star-forming regions in Orion. However, our traceback reveals several literature populations as old as 20 Myr that share a close configuration around the time of their formation, suggesting a multi-generational star-forming event like in the Local Bubble. For ease of reference, we refer to the collective of populations co-spatial or co-moving with these three low-mass populations as the ``Orion Family''.

TOR1B and AriS both have largely divergent velocities relative to the rest of the complex, with angles between their directions of motion and velocities relative to the Orion Family average of 23$^{\circ}$ and 35$^{\circ}$, respectively. Their motions are therefore consistent with material swept up in an expanding bubble. NGC 2232 and LP 2439 are older than both, as are some outlying generations in Orion OB1 \citep{Kerr21}, all of which serve as plausible feedback sources. VulE's connection to Orion is more uncertain due to its high velocity and old age, and we explore this in Appendix \ref{app:vule}.

The low-mass associations connected to Orion are relatively young, so we can assess their connections to present-day gas and dust structures in the Orion-Eridanus Superbubble. We find that AriS aligns with the eastern edge of the ``North Rim'' in \citet{Joubaud19}, with positions overlapping with the clouds MBM 16 and 17. These clouds have RVs of 7.4 and 7.9 km s$^{-1}$ respectively in \citet{Magnani85}, and overlap with a 9 km s$^{-1}$  H I RV component in \citet{Joubaud19}. AriS has a similar average RV of 11.0 km s$^{-1}$. This is slightly slower than the North Rim clouds relative to the center of the Superbubble, but some high-quality RVs in AriS overlap with their velocities. AriS therefore has velocities consistent with the Orion-Eridanus Superbubble, although its extension beyond the bubble edge and marginally slower velocities suggest that it may have formed early from lower-velocity gas. If bubble expansion stalled after the formation of AriS due to a lull in supernova activity, AriS could have overtaken the edge of the bubble before the bubble was accelerated again by new supernovae in Orion OB1.

TOR1B overlaps on-sky with part of the North Rim that hosts several clouds with varied distances and velocities \citep{Edenhofer24}. Its 289 pc distance and 19.8 km s$^{-1}$ RV do not align with specific \citet{Magnani85} gas clouds, but those properties are consistent with the range given for the North Rim in \citet{Joubaud19}. TOR1B's RV is on the upper end of the North Rim's 4-25 km s$^{-1}$ range, which is expected for a population in part of the bubble moving largely tangential to the sun compared to the blue-shifted material on the near edge. 

\subsubsection{Leo}

The Leo Association has no co-moving literature populations, so the only related populations were co-spatial when LeoC formed near the Galactic midplane. We show their velocities in $\xi$-$\zeta$ space (co-rotating $X$-$Z$ galactic coordinates) in Figure \ref{fig:othertraces}, where Leo passes three populations in the Orion Family near the time of LeoC's formation. LeoE's 22 Myr age from \citetalias{Kerr25b} is similar to the early generations in Orion and their velocities are convergent, so Leo is not related to bubble material from Orion. However, the velocities of LeoC and LeoE differ by $(\Delta U, \Delta V, \Delta W) = (-2.1, 0.9, -4.7)$ km s$^{-1}$ during their closest approach in traceback 20 Myr ago. Assuming that the average of all populations in the central component of the Orion Family (Orion OB1, MSW, and $\lambda$ Ori) represents local gas velocities, Leo's average relative velocity is $(U^{\prime},V^{\prime},W^{\prime}) = (5.8, -2.3, 26.6)$ km s$^{-1}$. The velocity difference between LeoE and LeoC almost directly opposes this vector, aligning within 1$^{\circ}$ in $\xi$-$\eta$ space and 12$^{\circ}$ in $\xi$-$\zeta$ space. This velocity difference is therefore consistent with deceleration relative to the Orion Complex. The $12^{\circ}$ deviation between the $\zeta$ and $\xi$ components ($\sim1$ km s$^{-1}$) may suggest additional acceleration in the Galactic plane, but this could be negated by our choice of rest frame. 

\section{Discussion} \label{sec:discussion}

\subsection{Coherence of Cluster Families} \label{sec:clfams_coher}

In Section \ref{sec:conn-lit}, we show that many of our low-mass associations share common formation sites with populations connected to nearby bubble structures like the Local Bubble and Orion-Eridanus Superbubble. However, truly connected star-forming events must share some chain of causality with the populations that come before them, be it through a triggering pattern of younger ages and higher velocities relative to a feedback source, or a collision between two known velocity components. To assess whether a single bubble-driven star forming event can explain the existence of these populations, we assemble models for the expansion of the Local and Orion-Eridanus Bubbles, and assess whether the velocities in potential bubble-tracing populations can accurately reproduce present-day bubble extents. 

\subsubsection{Model Assembly}

We assemble our models by assuming that bubble expansion is initiated by the oldest stellar generation in the parent family, and that the bubble expands spherically in the rest frame of the gas. We set the gas rest frame to the average across literature members of the Family, including more massive populations in our low-mass set like CaNMoS (which has similar demographics to Taurus), and excluding late-forming populations with velocities that may reflect recent acceleration by supernovae rather than initial gas velocities, such as CrA in the Local Family and $\sigma$ Ori and the ONC in Orion. 

When possible, we assume that young populations with outward-directed velocities relative to the bubble model trace the edge of that bubble. We use their velocities to set the model expansion rate, and update that expansion rate at the time a new potentially bubble-tracing population forms. When populations cannot constrain bubble expansion, we use the expansion rates of analogous populations. 

\subsubsection{Local Bubble} \label{sec:bubble-local}

\begin{figure*}
\centering
    \includegraphics[height=20cm]{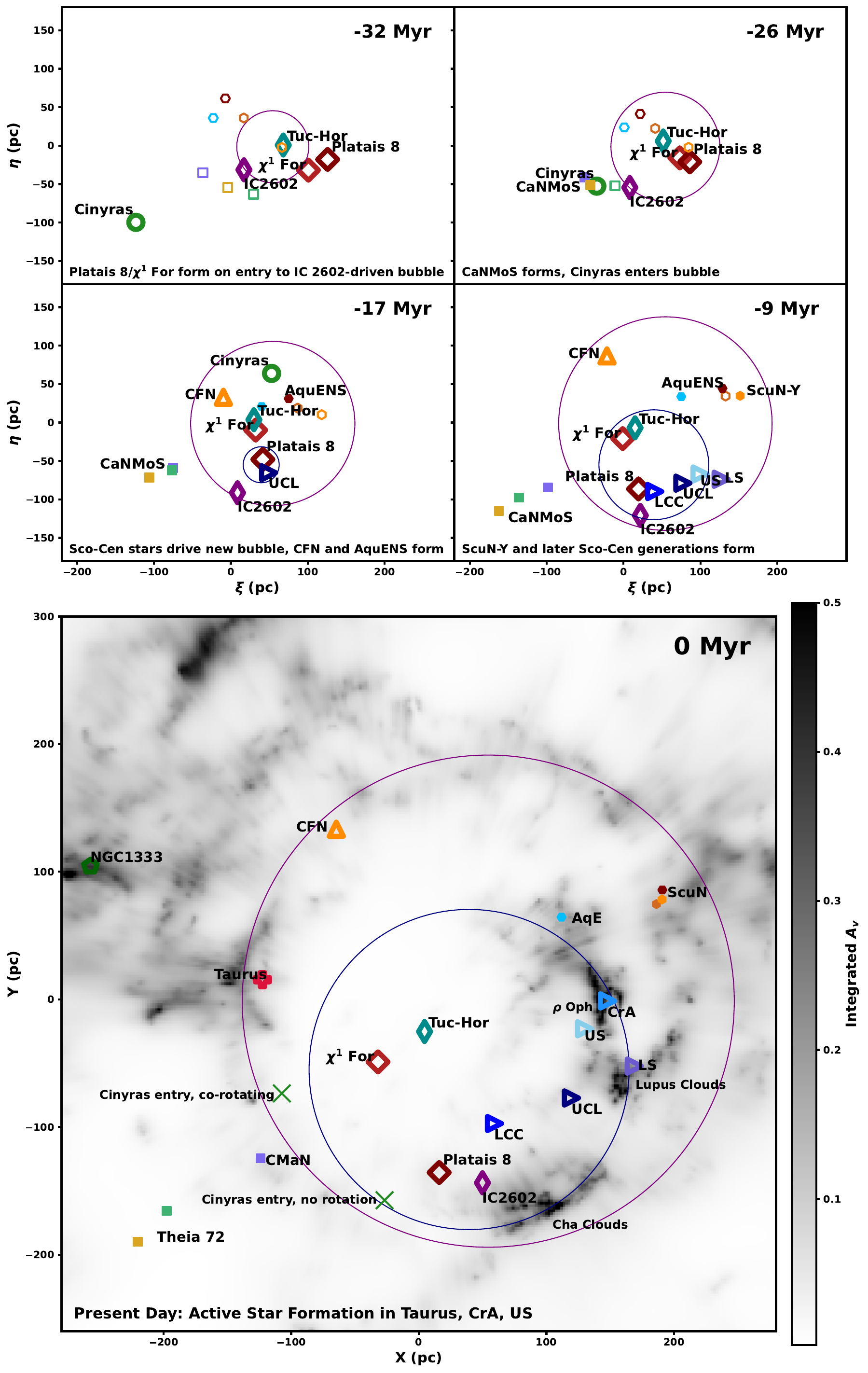}
\caption{Traceback of young associations connected to Local Family. In all panels, literature populations are shown as larger markers, and low-mass associations use smaller markers that are open before formation and filled after formation. Circles indicate the edges of our Local Bubble models, with the outer model driven by IC 2602 in purple, and the inner model driven by early generations in Sco-Cen in blue. In the bottom panel, we show all features in the present day overlaid on the \citet{Edenhofer24} dust map. The edges of the Local Bubble align closely with our models, with the inner model following clouds associated with Sco-Cen (labelled), and the outer model following Taurus and several other lower-density structures. We also mark the entry point of RSG 5 in Cinyras into the outer bubble model, which aligns closely with the center of the Mirzam Tunnel when assuming no bubble rotation rather than rotation with the galaxy. The online-only version can be interacted with in 3D by zooming, rotating, and panning. We also provide buttons to toggle the display of the dust, populations, and bubble models. Mousing over data points provides positions and population names. }
\label{fig:localbubble}
\end{figure*}

The oldest populations in the Local Family are IC 2602 and Tuc-Hor, which have similar ages of $\sim$ 45 Myr \citep{Kraus14, Randich18, Wood23}. IC 2602 is the more massive of the two with $\sim$600 members \citep{Liu25}, compared to $\sim$400 in Tuc-Hor \citep{Popinchalk23}, so we adopt IC 2602 as the bubble driver, with bubble expansion starting at its 43.7 Myr adopted age. These populations may be related, as proposed in \citet{Gagne21}, and filamentary structures with scales consistent with the 58 pc separation between them at formation are known to exist \citep{Jackson10}. However, these populations do not diverge at formation, so there is little indication that one traces the edge of a bubble driven by the other. We therefore adopt 4 km s$^{-1}$ as the expansion rate, following the velocity spread in the similarly-populated Circinus association \citep{Kerr25}. 

Four populations form later with outward-oriented velocities that may trace the Local Bubble: CaNMoS at $\sim$26 Myr, CFN-1 in CFN at 17.8 Myr, ScuN-Cen and -Halo in ScuN-Y at 6.8 and 12.8 Myr, respectively, and L1524 in Taurus at 1.6 Myr. At 20 Myr, when CaNMoS is well-enough separated from the bubble center to mitigate uncertainty in its placement, CMaN and Theia 72 have a speeds relative to the center of the bubble of 2.8 km s$^{-1}$ and 7.1 km s$^{-1}$, respectively. This velocity range overlaps with our 4 km s$^{-1}$ expansion rate, necessitating no expansion rate changes. CFN-1 also hosts outward velocities soon after formation consistent with that expansion speed. When L1524 and the last stellar generation in ScuN-Y form on near-opposite ends of the model bubble, both show significantly higher outward velocities of 8.1 and 6.3 km s$^{-1}$, respectively, suggesting a recent increase to the expansion velocities. We therefore adopt a 7 km s$^{-1}$ expansion speed for the last 6 Myr of our model. An earlier start to this higher expansion speed could be justified by ScuN-Halo's older age, but ScuN has few enough well-characterized stars to merit caution in its velocities and traced-back positions. 

The Sco-Cen Association is the largest stellar population that we group into the Local Family, and the velocities of its oldest components are largely tangential relative to the center of our bubble model at formation. It therefore likely does not trace the bubble edge. However, Sco-Cen has a known velocity spread of 6 km s$^{-1}$, with the youngest populations having the highest velocities. This was proposed by \citet{Posch25} to indicate feedback-driven sequential star formation, so Sco-Cen may have driven its own bubble that locally dominated over any larger bubble structure. We therefore model a separate bubble centered on UCL with a 6 km s$^{-1}$ expansion rate, using the same gas rest frame as for the main bubble model. 

Figure \ref{fig:localbubble} show the evolution of the bubble models alongside the stellar populations that we associate with it. All populations with outward velocities form within $\sim$50 pc of the edge of our bubble models, making them consistent with bubble-driven formation at a level consistent with typical asymmetries \citep{ONeill24}. Other populations have uncertain connections to our bubble model. Tuc-Hor may share an origin with the proposed bubble progenitor IC 2602, but its age and velocities are inconsistent with formation in bubble material. Platais 8 and $\chi^1$ For form $\sim$5 Myr before they enter our model bubble, so rather than forming from bubble material, they may originate in a collision with bubble material. All three of these populations remain inside our model bubble for most of their $30-45$ Myr lifetimes, so any supernovae there or in related populations like Carina and Columba could contribute to bubble expansion. Early generations in Sco-Cen also form near the edge of our bubble, and their orientation parallel to the bubble edge at formation supports \citetalias{Kerr21}'s idea that the spatially distributed and largely coeval old generations in Sco-Cen originate in a supernova shell-driven structure called the Libra-Centaurus Arc. The lack of outward-oriented velocities suggests that any star formation triggering did not substantially accelerate the cloud, likely due to the mass Sco-Cen's progenitor cloud, as Sco-Cen's membership of $1.3 \times 10^4$ stars is an order of magnitude larger than Tuc-Hor and IC 2602 combined \citep{Ratzenbock23}.

The bottom panel of Fig. \ref{fig:localbubble} shows our bubble models in the present day against the \citet{Edenhofer24} dust map. Our model replicates the locations of most overdensities associated with the Local Bubble, including an apparent nested bubble structure. Taurus and other lower-density clouds fall along the older IC 2602-driven bubble model, while clouds in Sco-Cen like Lupus and CrA follow the smaller Sco-Cen-driven nested bubble, which is open on the side opposite to the present-day Sco-Cen clouds. The outer bubble more closely follows the gas overdensities on the side opposite the dense material in Sco-Cen, suggesting that the wider bubble in that direction and possible recent acceleration in Taurus can be explained by feedback from supernovae in Sco-Cen escaping its bubble, ejecting remnant bubble material and directly exposing gas there to feedback. The asymmetric gas distribution in the inner bubble also suggests that early feedback sources in Sco-Cen were off-center, a configuration consistent with triggered star formation from a larger bubble on the side of the cloud toward the bubble center. The TW Hydrae Association has both outbound velocities relative to Sco-Cen and a position offset opposite to other Sco-Cen affiliated populations, so it may represent formation in the opposite edge of the inner bubble. However, its current position of $(X, Y, Z) = (6, -40, 23)$ is well inside that model bubble, requiring slower expansion than our model predicts \citep{MiretRoig25}. 

\subsubsection{Orion-Eridanus Superbubble}\label{Orion-Eridanus Family}

\begin{figure*}
\centering
    \includegraphics[height=21cm]{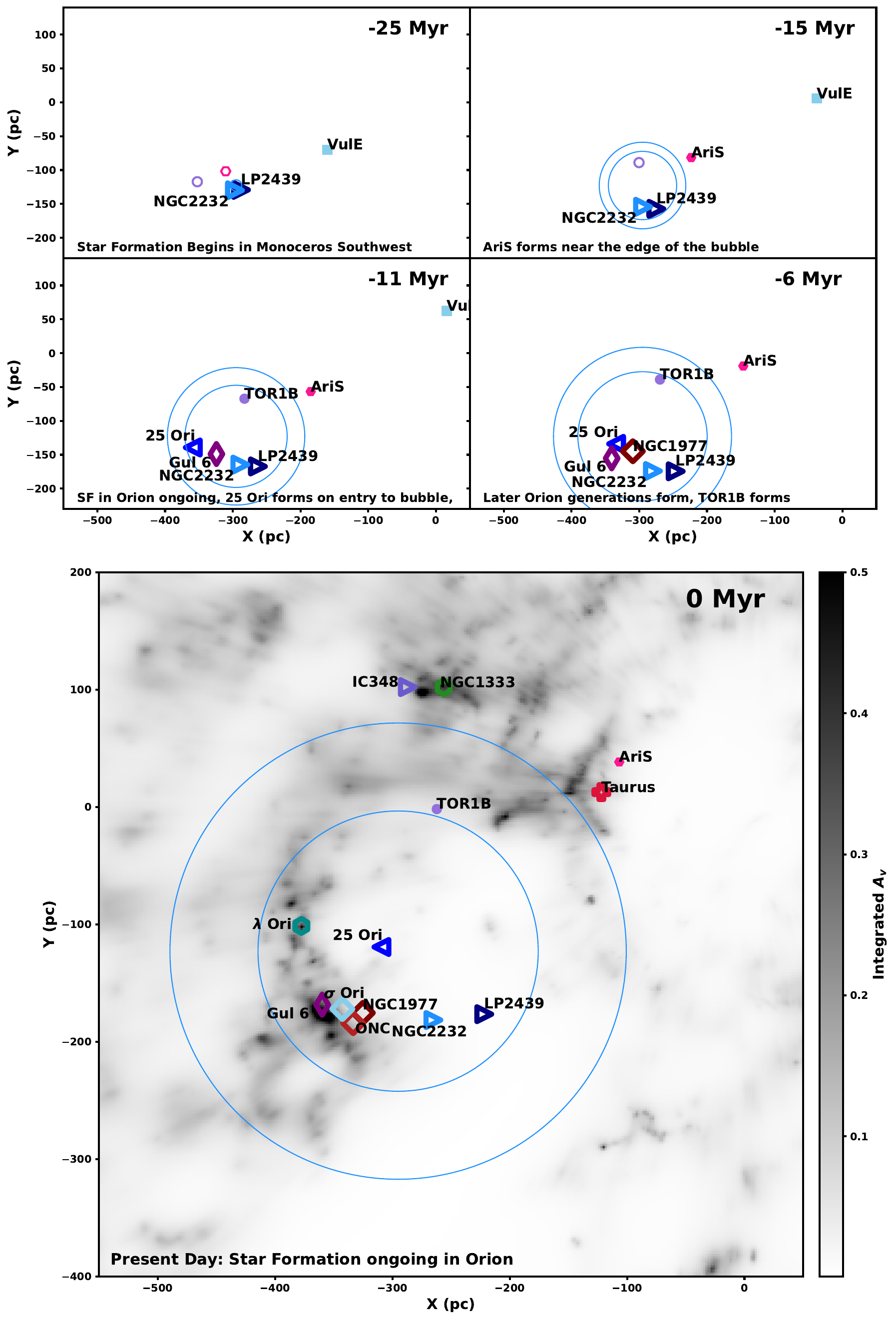}
\caption{Same as Fig. \ref{fig:localbubble}, but for the Orion Family. The two circles indicate the fast (large) and slow (small) expansion models, which capture the uncertainty and asymmetry in the Bubble's expansion velocity.}
\label{fig:orionbubble}
\end{figure*}

The earliest stellar generations in Orion-Eridanus that serve as potential bubble drivers are NGC 2232 and LP 2439 in MSW, which formed near-co-spatially. Their velocities differed by $\sim$2.5 km s$^{-1}$ at formation, and this was mostly divergent 15 Myr ago, when they were sufficiently separated to mitigate traceback uncertainties. Other subgroups in MSW have similar transverse velocity separations of $\sim3$ km s$^{-1}$ \citep{Kerr21}, consistent with known bubbles, so we take 3 km s$^{-1}$ as the initial expansion rate, selecting NGC 2232 as the bubble origin due to its slightly older 27.8 Myr age.

Most later-forming populations have outward-oriented velocities in $\xi$ vs $\eta$ space at formation, but the $\zeta$ components are often discrepant. AriS and TOR1B both have velocities within $\sim25^{\circ}$ of outward if the $\zeta$ direction is ignored, nearly half the angle seen when including it. Some literature populations follow this pattern like NGC 1977 and $\lambda$ Ori, which have angles relative to radial of $21^{\circ}$  and $12^{\circ}$ without $\zeta$, and $\sim55^{\circ}$ with it. These discrepancies appear to be driven by our choice of bubble origin, as a bubble center further below the galactic midplane would bring these velocities closer to being outward-oriented. Some early generations in Orion have high $-W$ velocities that bring them far further below the galactic midplane than MSW, such as the 21 $\pm$ 3 Myr old ASCC 20, a dynamically outlying component of Orion OB1a \citep{Kos19}. Orion is far more populated than MSW with just the relatively old components within 333 pc outnumbering MSW by nearly a factor of 3 \citep{Kerr21}. It is plausible that early generations in Orion became dominant over MSW soon after formation, moving the bubble center further from the galactic midplane and contributing to the Bubble's well-established asymmetry, which is also oriented largely out of the Galactic midplane \citep{Pon14}.

To account for uncertainties in the bubble-driving population and broader asymmetries, especially in the vertical $\zeta$ direction, we present two possible bubble radii starting at $\sim$20 Myr: one assuming that only radial components of motion reflect expansion, and one taking any outward-directed velocity vector in $\xi$ vs $\eta$ space to have purely outward motions that trace bubble expansion. Starting at 20 Myr, we set the bubble expansion speed to 6 km s$^{-1}$ in the slow model and 9 km s$^{-1}$ in the fast model, following AriS's velocities. We reduce the fast model to 6 km s$^{-1}$ and slow model to 4 km s$^{-1}$ at 12 Myr, following the velocities of Gulliver 6. After this, NGC 1977 and TOR1B form within 1 Myr of each other with $\sim$6 km s$^{-1}$ total velocities and outward velocities of 3.7 and 5.0 km s$^{-1}$, consistent with the velocities in Gulliver 6. Similar velocities continue in $\sigma$ Ori 3 Myr before present. The ONC's velocities are slower, but the 10 km s$^{-1}$ $v_{LSR}$ difference across its parent Integral-Shaped Filament \citep{Grossschedl21} suggests rapid acceleration, potentially by feedback from recent star formation or a collision with the bubble edge. We therefore keep the 4 km s$^{-1}$ expansion velocity in the slow model, but increase it to 15 km s$^{-1}$ for the last 3 Myr in the fast model to agree with present-day expansion velocities in the Orion-Eridanus Superbubble \citep{Reynolds79,Joubaud19}.

We show our final model in Figure \ref{fig:orionbubble}. In the present day, the slow bubble roughly follows a clear cavity in the \citet{Edenhofer24} dust maps, while the fast bubble approximates the current maximum extent of the Orion-Eridanus Superbubble \citep{Pon14}. The frequent velocity changes in our model may reflect different generations of stellar feedback that produced the nested shells identified by \citet{Oschendorf15}. Most populations form with positions and velocities consistent with material swept up in the bubble, but 25 Ori forms on entry to the bubble, similar to $\chi^1$ For and Platais 8 in the Local Family. IC 348 and NGC 1333 are near the edge of the fast bubble model radius in the present day, but \citetalias{Kerr21} shows earlier generations with ages of $\sim17$ Myr. These clusters were well beyond even the fast model's 50 pc radius when they formed, so it is unlikely that this bubble drove their formation.  

\subsubsection{Summary}

Our models show that the formation of most stellar populations in the Local and Orion families can be explained with expanding bubble models originating in their oldest populations, expanding with rates matching the velocities of younger populations with outward-oriented velocities. Most populations in these families form with positions and velocities consistent with swept-up bubble material. In some other cases, star formation may be more bubble-triggered than bubble-driven, as structures like 25 Ori, $\chi^1$ For, and Sco-Cen form on the edges of the bubble model, but lack velocities consistent with bubble expansion. Orion also shows significant asymmetry and a somewhat bimodal space-velocity offset between most of Orion OB1 and MSW, consistent with a cloud-cloud collision \citep{Wu17}. 

The scenarios we present here, where IC 2602 (43.7 Myr) and NGC 2232 (27.8 Myr) are the initial bubble-drivers in the Local and Orion Families, respectively, are among many possible scenarios. Not all velocities in later generations align perfectly with our predicted bubble velocities, but asymmetry is a known feature of both Bubbles, so this would be true for any spherical expansion model \citep{Cox87, Pon14, ONeill24}. Previous work has suggested much more rapid expansion timescales such as the 14 Myr old UCL/LCC-driven Local bubble proposed in \citet{Zucker22b}, and the 8 Myr old Orion-Eridanus bubble age from \citet{Brown94}. While our 43.7 and 27.8 Myr old bubble models in the Local and Orion-Eridanus Bubbles are not unique solutions, they show that an older bubble model can reproduce the velocities of most populations in their families. 

\subsection{The Mirzam Tunnel}

Cinyras and SCYA-79 pass the Local Family soon after its formation, with RSG 5 entering our model Local Bubble at 24 Myr, only 4 Myr after the youngest stellar generation in Cinyras formed \citep{Kerr24}. Based on the \citet{Chabrier05} IMF, the 4000-member Cinyras association should form with approximately 10 O stars, and their short ($<10$ Myr) lifetimes make a supernova during Cinyras' entry into the Local Bubble probable. We mark RSG 5's entry point into the Local Bubble in the Figure \ref{fig:localbubble}, showing its present-day site assuming both rotation that follows Galactic rotation (no rotation in $\xi$ vs $\eta$ space), and no rotation, which produces counter-clockwise rotation in the co-rotating frame similar to the trajectories in Fig. \ref{fig:canmostraceback}. Both potential entry points align with the broad gap on the lower-left edge of the bubble, suggesting that supernovae or feedback from OB stars in Cinyras provides a plausible explanation for the existence of the Mirzam Tunnel. 

CaNMoS may also produce a plausible explanation, as CMaN in particular spends most of its lifetime on the edge of the Local Bubble in the direction of the Mirzam Tunnel. However, it has an order of magnitude fewer members than Cinyras and is therefore expected to only have around 1 O star. It may therefore not have been capable of substantially affecting the bubble's morphology. Cinyras, in turn, is massive enough that even if no supernovae occured during its passage into the local bubble, it may have hosted material in its own bubble that collided with the Local Bubble. The no-rotation model places Cinyras' entry point into the Local Bubble at $(l, b) = (260, -25)$ in galactic coordinates, which is near the center of the tunnel extent shown in Figure 10 of \citet{ONeill24}. We therefore conclude that the entry of Cinyras into the Local Bubble is the most plausible explanation for the Mirzam Tunnel.

\subsection{Leo Association and Intermediate Velocity Clouds}

In Section \ref{sec:Leonode}, we reveal the highly vertical velocities in the Leo Association relative to the Galactic plane. No known young associations share similar velocities, but high vertical velocities like these are common among H I gas clouds in the Galactic halo. Much of this gas is thought to originate in the Galactic fountain, a process by which feedback-accelerated gas left behind after star formation is accelerated out of the Galactic plane prior to falling back in under the influence of gravity \citep{Wakker97, Marasco22}. Other origins have also been proposed, such as stripped gas from dwarf galaxies and dark matter sub-halos \citep{Lehner22}. Leo's largely vertical 20-25 km s$^{-1}$ velocities would not meet the definition of an intermediate velocity cloud (IVC) used in \citet{Lehner22} ($40$ km s$^{-1}$ $< |v_{LSR}| < 90$ km s$^{-1}$), but lower velocity clouds are known, and \citet{Rohser16} uses a broader definition that would include the Leo's parent cloud ($20$ km s$^{-1}$ $< |v_{LSR}| < 100$ km s$^{-1}$). 

IVCs are normally quiescent, non-star forming clouds \citep{Rohser14}. However, \citet{Rohser16} shows that under the influence of ram pressure, an atomic IVC can become a molecular cloud on the timescale of 1 Myr or less. Our observations in Leo may provide the first observational evidence of this. Not only does Leo have velocities consistent with known IVCs, but it also shows deceleration between the first (LeoE) and second (LeoC) generations, consistent with the effects of ram pressure. The extensive gas near Leo in the Orion Family at formation provides a plausible pressure source that may have triggered Leo's formation. 

IVCs usually have metallicities similar to gas in the Galactic plane, which supports their origin in Galactic Fountain material ejected by star formation in the midplane \citep{Lehner22}. However, variations exist, and some clouds in the Galactic halo, particularly those classified as HVCs, show sub-solar metallicities \citep{Tripp12, Fox16}. Leo's metallicity could therefore reveal the material that it formed out of, be it recent fountain material or more ancient halo gas. 

\subsection{Diversity of Fine Substructure} \label{sec:disc_finess}

The fine substructure introduced in Section \ref{sec:finess} and expanded on in Appendix \ref{app:microclusts} shows several qualitatively distinct varieties of substructure that may reflect different formation conditions. Both ScuN and AndS \added{show} small central substructures with velocities that largely match the enclosing structure, but younger ages that differ by $\sim26$ Myr in AndS and $\sim6$ Myr in ScuN (specifically ScuN-Y). This indicates surprisingly long-lived star formation in these low-mass associations, \added{and if the more tentative substructure in AndS is confirmed, its age difference relative to the rest of AndS would be }longer than Sco-Cen's entire accepted lifespan \citep{Kerr21, Ratzenbock23}. This longevity may be possible precisely because of their low masses, as a lack of supernovae limits the dispersal of gas clumps, allowing the cloud to survive the first generation and produce a new population after a period of dormancy. Similar old and young components exist in places like Perseus OB2, but these have a spatial offset between them, which is expected for any cloud subject to interactions with other clouds or feedback sources \citep{Kerr21}. It is therefore unclear whether a cloud can last 26 Myr without significant velocity changes, a question that simulations could help answer. 

The \added{robustly defined OphSE-B population} has a velocity outbound from the main component of OphSE on-sky, indicating that it may have formed from material accelerated away from the older generation. However, it lacks sufficiently precise RVs to assess this in 3 dimensions. The 2.6 km s$^{-1}$ velocity difference and 13.8 pc on-sky separation between the two components implies that divergence started only $\sim5$ Myr ago, \added{shortly before OphSE-B began forming}. Some other stars in OphSE's ``tail'' see similar divergence timescales despite their older ages, suggesting that these stars were ejected around the same time OphSE-B formed. A supernova could explain both features, accelerating OphSE-B's progenitor cloud and triggering star formation there while ejecting adjacent members, such as companions to the supernova progenitor. OphSE's 127-member population predicts $\sim0.4$ O stars given the \citet{Chabrier05} IMF, indicating that the presence of an early supernova-bearing star is plausible, but not necessarily expected. At $(X, Y, Z) = (208, 16, 45)$, the center of OphSE lies in a local minimum in the dust map (see Fig. \ref{fig:localbubble}), but the complex region makes it difficult to attribute any specific feature to a supernova. 

TOR1B hosts spatially separated clumps with similar velocities and ages that, while different, are similar enough that a change in outlier rejection could bring them into close agreement. \added{This substructure is therefore preliminary (see Sec \ref{sec:tor1b_ss}), but if membership refinement confirms the distinctness of these clumps at formation}, it would represent largely parallel formation \added{with a potential time offset, consistent with the} collapse of a filament toward two gas collection nodes \citep[e.g., see][]{Kirk13, Krause20}. 

\subsection{The Ubiquity of Low-Mass Associations}

The low-mass populations in \citetalias{Kerr25b} typically have masses exceeding $\sim20-25$ M$_{\odot}$, with the exception of LeoC and TOR1B, which both have $16.2 \pm 1.6$ M$_{\odot}$. Both outliers were found under near-ideal conditions: LeoC was far from contaminants in the Galactic midplane, and TOR1B shared position-transverse velocity space with the much larger 80 M$_{\odot}$ TOR1A. This implies that populations like ScuN-Y and TOR1B would not have been detected if not for chance alignments with more massive populations. Even if we assume that our sample is complete for $|Z|>100$~pc, the 15 populations in \citetalias{Kerr25b} would account for just over 0.1\% of the volume of the innermost 1 kpc, without considering their velocity extents. The discovery of two new associations with distinct origins in this tiny volume implies that the solar neighborhood could contain three orders of magnitude more populations like these. This estimate should be treated with caution due to the small number statistics and anisotropies in the solar neighborhood, but even increasing the sample of low-mass populations from 2 to 20 in a family like Orion would profoundly improve our ability to trace its gas evolution. 

To provide a theoretical estimate of how common low-mass populations should be in the presence of stellar feedback, we consider simulations from the STARFORGE suite, which simulate star formation from collapse to dispersal. The Fiducial 1 simulation, as shown in Figure 14 of \citetalias{Kerr25}, shows several feedback-accelerated clouds that produce 1-10 stars, making them potential analogs to groups like OphSE-B. \citetalias{Kerr25} applied the HDBSCAN clustering algorithm to this simulation, identifying 9 young associations that form at least 1 Myr after the expansion of the parent association commences, which we take to represent formation in the feedback-driven expansion stage. Only one population has $M>25$ M$_{\odot}$ in the initial HDBSCAN definition, while that number would increase to 4 after the assignment of outliers to the nearest HDBSCAN population in that work. However, those extended definitions include contaminants from older populations, which appear to be less prominent in our sample given the larger spatial and temporal separations between different populations. We therefore take the initial HDBSCAN definitions as more representative. Applying HDBSCAN to the four remaining STARFORGE Fiducial simulations reveals that 88\% of populations that form in the expansion stage have $M<25$ M$_{\odot}$ (36 of 41). This implies that populations with $M<25$ M$_{\odot}$ are about an order of magnitude more common than the populations discussed in this paper. Their mass contribution is small, representing only 3\% of the total stellar mass and 6\% of the stars. These simulations also show that many dynamically distinct gas clouds need not form anything recognizable as a substructure, with some clouds producing individual stars. This suggests that even HDBSCAN-defined groups we discuss here with $M>1.2$~M$_{\odot}$ miss the stellar remnants of entire bubble-tracing gas clouds. 

\section{Conclusions} \label{sec:conclusion}

We have analyzed the origins of 16 low-mass associations in the solar neighborhood. Using positions, velocities, and ages across a suite of newly-characterized low-mass associations and large literature populations, we have performed dynamical traceback to identify common origins from which low-mass associations emerge. We reveal that most populations form in common structures with both other low-mass associations and larger complexes, tracing a range of features from feedback-driven bubbles to intermediate velocity clouds. Our key findings are as follows:

\begin{enumerate}
    \item Three pairs of low-mass populations originate in common formation sites, comprising three new substructured associations: CaNMoS (containing Canis Major North and Theia 72), AquENS (Aquila East and the ScuN-AE subcomponent in Scutum North), and Leo (LeoC and LeoE).
    \item CaNMoS, ScuN-Y, and AquENS have motions consistent with an origin in the Local Bubble alongside the rest of what we call the Local Family. Using probable feedback sources to define a bubble model, we show that the Local Bubble's asymmetric and potentially nested structure is consistent with a contiguous star-forming event originating in IC 2602 $\sim$44 Myr ago, with structures forming through a combination of swept-up expanding bubble material and collisions with other clouds.   
    \item TOR1B and AriS formed co-spatially with populations connected to the Orion Complex, and host velocities consistent with the edge of the Orion-Eridanus Superbubble. Like in the Local Bubble, the $\sim28$ Myr history of populations we connect to the ``Orion Family'' can reproduce the current extent of the Orion-Eridanus Superbubble. 
    \item The location of the Mirzam Tunnel is consistent with a collision between the Cinyras complex and our long-lived Local Bubble model. 
    \item The Leo Association represents the first young association with a probable origin in an intermediate velocity cloud. It exhibits both highly vertical velocities of $>20$ km s$^{-1}$, and a deceleration signature between two time-separated generations consistent with the effects of ram pressure interactions with material in the Galactic midplane, especially in the Orion Complex. 
    \item Only SCYA-54 and CasE reside in well-populated sections of the literature sample but lack clear relationships to those populations, indicating that most low-mass associations form out of larger bubble structures or complexes
    \item Four populations host small substructures with masses as low as 1.6 $M_{\odot}$\added{, with the substructure in ScuN and OphSE being robustly defined}. They appear to trace distinctly different processes, from undispersed gas in ScuN and AndS, to supernova-accelerated gas in OphSE, to multiple gas collection nodes in a collapsing filament in TOR1B. 
\end{enumerate}

Our results show that low-mass associations often emerge as byproducts of larger star-forming events, tracing features from the edges of expanding bubbles to collisions between IVCs and dense clouds in the Galactic midplane. The discovery of progressively lower-mass structures, including the coherent substructures in ScuN, AndS, OphSE, and TOR1B, suggests that most low-mass populations remain undiscovered. These structures may provide an invaluable record of recent star formation, tracing the evolution of lower-mass gas structures embedded in larger structures that cannot produce massive associations or clusters. 




\begin{acknowledgments}
The Dunlap Institute is funded through an endowment established by the David Dunlap family and the University of Toronto. RMPK received funding from the Heising-Simons Foundation. J.C. acknowledges support from the Agencia Nacional de Investigaci\'on y Desarrollo (ANID) via Proyecto Fondecyt Regular 1231345, and by ANID BASAL project FB210003. J.G.F-T gratefully acknowledges the grants support provided by ANID Fondecyt Postdoc No. 3230001 (Sponsoring researcher), the Joint Committee ESO-Government of Chile under the agreement 2023 ORP 062/2023, and the support of the Doctoral Program in Artificial Intelligence, DISC-UCN. This paper made use of new data from the Sloan Digital Sky Survey V (SDSS-V). Funding for SDSS-V has been provided by the Alfred P. Sloan Foundation, the Heising-Simons Foundation, the National Science Foundation, and the Participating Institutions. SDSS acknowledges support and resources from the Center for High-Performance Computing at the University of Utah. SDSS telescopes are located at Apache Point Observatory, funded by the Astrophysical Research Consortium and operated by New Mexico State University, and at Las Campanas Observatory, operated by the Carnegie Institution for Science. The SDSS web site is \url{www.sdss.org}.

SDSS is managed by the Astrophysical Research Consortium for the Participating Institutions of the SDSS Collaboration, including the Carnegie Institution for Science, Chilean National Time Allocation Committee (CNTAC) ratified researchers, Caltech, the Gotham Participation Group, Harvard University, Heidelberg University, The Flatiron Institute, The Johns Hopkins University, L'Ecole polytechnique f\'{e}d\'{e}rale de Lausanne (EPFL), Leibniz-Institut f\"{u}r Astrophysik Potsdam (AIP), Max-Planck-Institut f\"{u}r Astronomie (MPIA Heidelberg), Max-Planck-Institut f\"{u}r Extraterrestrische Physik (MPE), Nanjing University, National Astronomical Observatories of China (NAOC), New Mexico State University, The Ohio State University, Pennsylvania State University, Smithsonian Astrophysical Observatory, Space Telescope Science Institute (STScI), the Stellar Astrophysics Participation Group, Universidad Nacional Aut\'{o}noma de M\'{e}xico, University of Arizona, University of Colorado Boulder, University of Illinois at Urbana-Champaign, University of Toronto, University of Utah, University of Virginia, Yale University, and Yunnan University.
\end{acknowledgments}

\begin{contribution}

This paper was led by, written by, and submitted by RMPK. ALK supervised RMPK's graduate work, during which the observations that preceded this work were completed. ALK and all other co-authors contributed by reviewing the completed work and recommending changes. 


\end{contribution}

%
\facilities{\textit{Gaia}, Smith, Sloan, DuPont}

\software{astropy \citep{Astropy13, Astropy18, astropy22},  numpy \citep{numpy}, pandas \citep{pandas}, LMFIT \citep{lmfit}
          saphires \citep{Tofflemire19}, matplotlib \citep{Hunter07}}

\appendix

\section{Fine Substructures} \label{app:microclusts}

\subsection{Scutum North} \label{app:scunss}

\begin{figure*}
\centering
    \includegraphics[width=18cm]{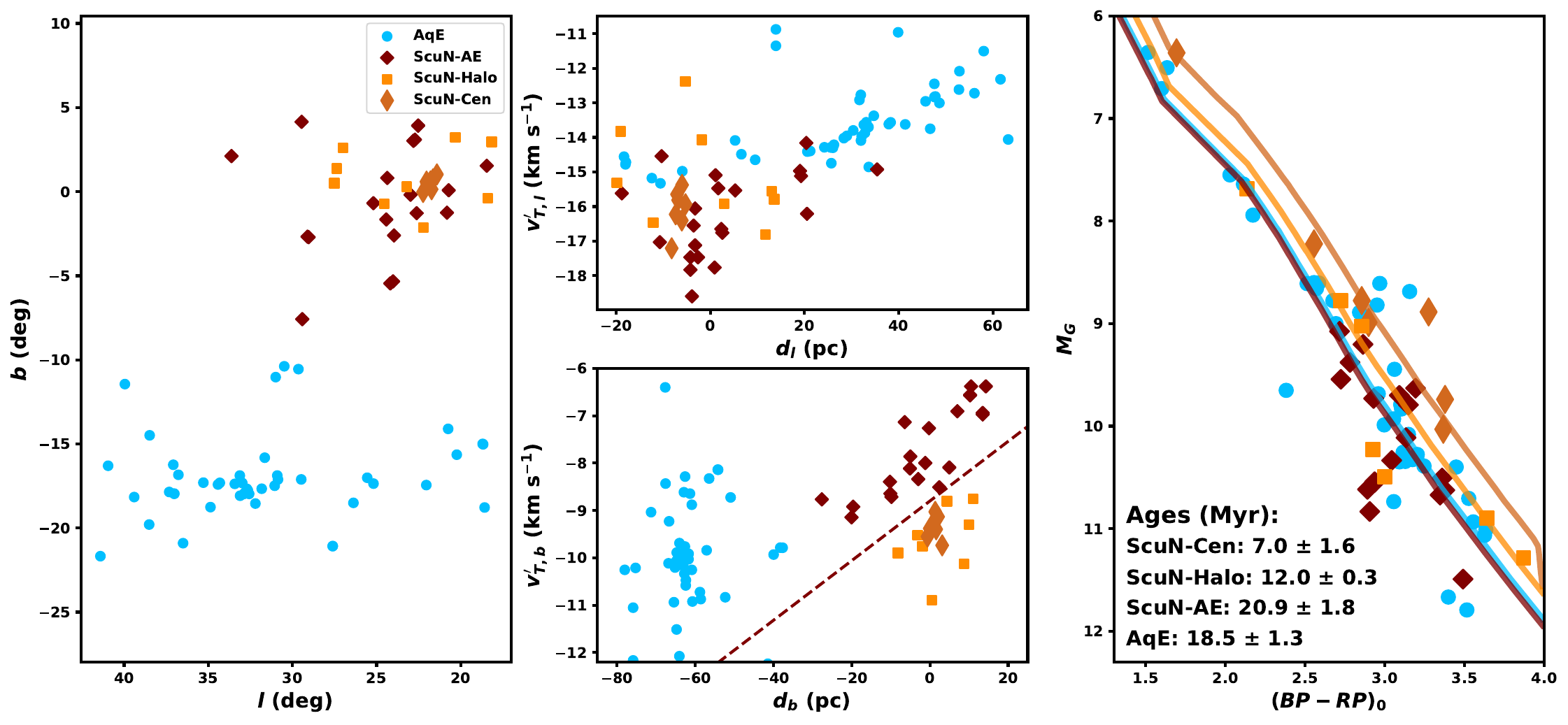}
\caption{Substructures in ScuN, with AqE for reference. We show $l$ vs $b$ spatial coordinates at left. The expansion-corrected transverse velocity distribution in $l$ and $b$ are shown in the middle column. All three components are separated using some combination of space and velocity cuts. ScuN-Cen was identified by isolating a central clump, and the remaining stars were divided further with a linear cut in $d_b$ vs $v_{T,b}$ parameter space (indicated by the dashed line), splitting it into a larger halo around ScuN-Cen (ScuN-Halo), and a component that aligns with expansion from AqE (ScuN-AE). All three components are distinct in the CMD (right), where members of ScuN-Cen are consistently the youngest, followed by ScuN-Halo (with two outliers), and then ScuN-AE, which is consistent with AqE. The ages of subgroups are annotated, and the corresponding best-fit isochrones are shown with colors matching the markers.}
\label{fig:scunsc}
\end{figure*}

At (RA, Dec) = (277.5, -9.5), ScuN has a compact stellar clump that we separate with $276<\text{RA}<279$ and $-11<\text{Dec}<-8$. All but three spatially outlying stars that pass this cut also have $-10<v_{T,b}<-8.5$ km s$^{-1}$, which we impose on the remaining stars. We refer to this clump as ScuN-Cen, and show its location in space and velocity coordinates in Figure \ref{fig:scunsc}. There we also provide a CMD of ScuN and AqE, showing that nearly all stars in ScuN-Cen are photometrically younger than the stars outside of it. 

To \added{assess the age distinctness of ScuN-Cen and other populations, we compute ages, largely following the method from \citetalias{Kerr25b}, but with changes to optimize for cases with limited stars. To preserve as many probable young members as possible, we remove any restrictions on $P_{spatial}$, and mitigate contaminants by requiring either $P_{fin}>0.9$ or $P_{Age<50 Myr}>0.05$. The former restriction ensures that all stars with robust membership markers are included, while the latter roughly includes stars with a $>$50\% membership chance for a star with $P_{spatial}< 0.1$, covering a wider range of photometrically young stars that reduces the risk of a young bias due to the removal of stars on the lower edge of the PMS. We also remove the $M<0.2$M$_{\odot}$ cut from \citetalias{Kerr25b}, prioritizing sample size over the low-mass model discrepancy that this cut alleviates. We take $10^4$ bootstrapped samples of the remaining stars, permitting repeat selections, and take the median and standard deviation as the age and uncertainty. We only clip the samples at 2-$\sigma$ as in \citetalias{Kerr25b} for populations with at least 5 surviving PMS stars to avoid over-clipping. Using this method,} we find that ScuN-Cen has an age of 7.0$\pm$1.6 Myr, $\sim$3 Myr younger than ScuN's bulk age and more than 10 Myr younger than AqE.

\begin{deluxetable*}{cccccccccccccccccc}
\tablecolumns{18}
\tablewidth{0pt}
\tabletypesize{\scriptsize}
\tablecaption{Demographics and mean properties of new substructure in AndS, OphSE, ScuN, and TOR1B. \added{We classify the substructures in ScuN and OphSE as robust ($<10\%$ false positive rate), while AndS-Cen is tentative. TOR1B requires membership refinement before its robustness can be properly assessed, while membership refinement will also be necessary to resolve cross-contamination between ScuN-AE and ScuN-Halo. }}
\label{tab:micclus}
\tablehead{
\colhead{Group} &
\colhead{N\tablenotemark{a}} &
\colhead{M\tablenotemark{a}} &
\colhead{RA} &
\colhead{Dec} &
\colhead{d} &
\colhead{$\overline{\mu_{RA}}$} &
\colhead{$\overline{\mu_{Dec}}$} &
\colhead{$\overline{v_{T,l}}$} &  
\colhead{$\overline{v_{T,b}}$} &  
\colhead{$\overline{v_r}$} &  
\colhead{$R_{hm}$} &  
\colhead{$\sigma_{1D}$} &  
\colhead{$\sigma_{vir}$} &
\multicolumn{2}{c}{Vir. Rat.} &
\multicolumn{2}{c}{Age}   \\
\colhead{} &
\colhead{} &
\colhead{(M$_{\odot}$)} &
\multicolumn{2}{c}{(deg)} &
\colhead{(pc)} &
\multicolumn{2}{c}{(mas yr$^{-1}$)} &
\multicolumn{3}{c}{(km s$^{-1}$)} &
\colhead{(pc)} &
\multicolumn{2}{c}{(km s$^{-1}$)} &
\colhead{val}&
\colhead{err}&
\colhead{val}&
\colhead{err}}
\startdata
AndS-Cen & 5 & 1.6 & 5.7 & 30.9 & 166.0 & 14.6 & -3.0 & 11.1 & -3.9 & 9.1 & 0.82 & 0.054 & 0.041 & 3.0 & 1.3 & 17.1 & 1.5 \\
OphSE-B & 7 & 1.6 & 259.0 & -20.9 & 230.0 & -3.2 & -12.9 & -13.6 & -5.1 & -11.1 & 0.3 & 0.136 & 0.069 & 3.3 & 1.5 & 2.1 & 1.4 \\
ScuN-Cen & 11 & 3.4 & 277.2 & -9.6 & 207.0 & 0.6 & -19.2 & -16.4 & -9.3 & -7.3 & 1.37 & 0.176 & 0.046 & 12.6 & 4.7 & 7.0 & 1.6 \\
ScuN-Halo & 18 & 7.2 & 277.4 & -7.7 & 205.0 & 1.3 & -18.8 & -15.5 & -9.5 & -9.2 & 8.15 & 0.827 & 0.028 & 21.1 & 8.5 & 12.0 & 0.3 \\
ScuN-AE & 32 & 12.2 & 279.2 & -7.9 & 204.0 & -0.4 & -18.4 & -16.0 & -7.8 & -8.3 & 13.65 & 0.945 & 0.028 & 21.0 & 8.5 & 20.9 & 1.8 \\
TOR1B-A & 14 & 5.2 & 65.6 & 15.3 & 286.0 & 3.1 & -3.3 & 6.1 & 0.4 & 20.3 & 0.99 & 0.485 & 0.067 & 4.8 & 1.7 & 4.5 & 0.4 \\
TOR1B-B & 6 & 4.3 & 67.2 & 14.8 & 304.0 & 3.0 & -3.6 & 6.7 & 0.2 & 20.4 & 1.47 & 0.268 & 0.05 & 6.4 & 2.4 & 11.4 & 1.2 \\
\enddata
\tablenotetext{a}{For our mass and stellar population measurements, we adopt a 10\% systematic uncertainty, following \citetalias{Kerr25b}.}
\tablenotetext{b}{Only one star used, uncertainty = 4 km s$^{-1}$ (\textit{Gaia})}
\vspace*{0.1in}
\end{deluxetable*}

The rest of ScuN can be split in $d_b$ vs $v_{T,b}^{\prime}$ distance vs expansion-corrected transverse velocity space, revealing a component centered on ScuN-Cen (ScuN-Halo), and a linear component with a slope that places it along AqE's expansion trend (ScuN-AE). The dynamical distinction between ScuN-Halo and ScuN-AE is corroborated by their ages, with ScuN-Halo at $12.0\pm0.3$ Myr, and ScuN-AE at $20.9\pm1.8$ Myr. The latter is consistent with AqE's $18.5\pm1.3$ Myr age, supporting the status of ScuN-AE as an outlying component of AqE. ScuN-Cen and ScuN-Halo therefore have ages that differ from each other and ScuN-AE by 5 Myr or more, meeting our age distinctness cut. Using the demographics routine from \citetalias{Kerr25b}, we calculate bulk properties for the subgroups, and report the results in Table \ref{tab:micclus}. 

\begin{figure*}
\centering
    \includegraphics[width=17.5cm]{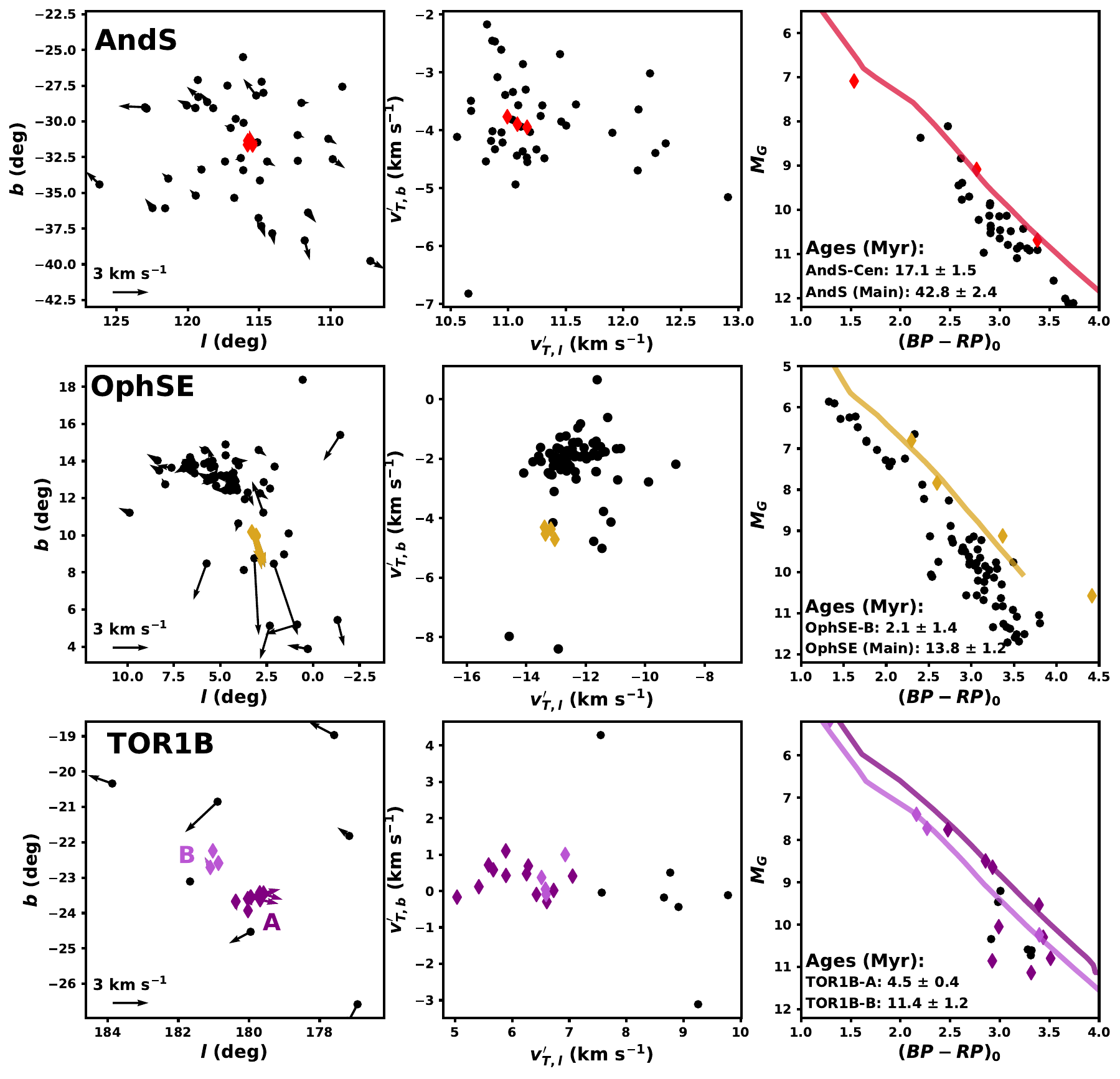}
\caption{Substructures in AndS, OphSE, and TOR1B, with the association annotated by row. Stars outside the substructures are shown with black circles, and the dynamically and temporally distinct subcomponents are marked with colored diamonds. The left column shows the on-sky spatial distribution in $l$ vs $b$ coordinates, with arrows that indicate virtual expansion-corrected transverse velocity relative to the population's average, and the middle column shows $l$ vs $b$ transverse velocity. The right column shows a CMD for each population, revealing that all substructures have distinct ages, especially OphSE-B in OphSE and AndS-Cen in AndS. Their ages are annotated, and the best-fit isochrones are shown with colors matching the markers.}
\label{fig:ophse}
\end{figure*}

To \added{assess robustness, we randomly select N stars from each substructure $10^5$ times, where N is the number of valid PMS stars in the population being compared against, and compute the fraction of samples that produce ages consistent with the other groups. We define our false positive rate as the chance of a population distinct at this level being identified by chance across our sample of 16 low-mass populations, assuming that all groups have one chance of a feature like this emerging. ScuN-Cen's stars are all photometrically younger than any stars in ScuN-Halo, so there is no overlap in their age samples. With 6 PMS stars in each, the probability of all stars in ScuN-Cen being photometrically younger by chance is 0.1\%, with a 2\% chance of this happening somewhere in our low-mass sample. We therefore consider ScuN-Cen's status as an age-differentiated substructure within ScuN-Halo to be robust, even if the groups only marginally meet our distinctness criterion. }

For \added{ScuN-AE, 1.1\% of samples are below the upper age limit of ScuN-Halo, and 0.7\% are below the upper age limit of ScuN-Cen, corresponding to a 17\% and 10\% chance, respectively, of this occurring somewhere in our 16-population low-mass sample. Due to these chances exceeding 10\%, we consider ScuN-AE's age differentiation tentative. However, the positions of non-ScuN-AE members of ScuN span more than 100 pc from ScuN-AE's median at formation in Fig. \ref{fig:combpops}, with positional uncertainties at that time of no more than 20 pc. This strongly indicates that components of ScuN originate in different clouds. While the traceback in Section \ref{sec:results} suggests that ScuN-AE is dynamically related to AqE and the others are related to the Local Bubble, the large traceback spreads in Fig. \ref{fig:combpops} indicate that our method of splitting them is imperfect, and that they likely overlap in space-transverse velocity coordinates. The lack of age differentiation is therefore likely caused by cross-contamination, so we consider ScuN-AE to be a robust population with a tentatively defined extent.}

\subsection{Andromeda South}

Like ScuN, AndS hosts a dense clump of five high-probability members near its center, which we isolate by selecting stars within 3 pc of the cluster median. Most of these are central within the velocity distribution, as shown in Figure \ref{fig:ophse}, with two outliers that we clip with $v_{T,l}<11.4$. The remaining stars we select, hereafter AndS-Cen, are consistently high on the pre-main sequence (PMS), with an age of $17.1 \pm 1.5$ Myr. This is more than 20 Myr younger than the population's $42.8 \pm 2.4$ Myr bulk age from \citetalias{Kerr25b}, \added{which would make it distinct in age despite near identical bulk velocities compared to the rest of AndS}. With only two PMS members, this result is vulnerable to outliers such as unresolved binaries, although none of these stars have evidence for binarity. \added{Taking $10^5$ random samples of two stars from AndS outside AndS-Cen, we find that 1.3\% reproduce an age as young as AndS-Cen's upper age limit, corresponding to a 18\% chance of this happening somewhere in our sample. We therefore consider this structure tentative. }

\subsection{Ophiuchus Southeast} \label{app:osess}

OphSE is dominated by a substantial expanding core surrounded by a larger halo that extends into an apparent tail structure to the galactic southwest. Within this tail is a compact stellar overdensity containing four high-confidence members that forms a dense clump in both space (half mass radius $R_{hm}=0.3$ pc) and velocity coordinates ($\sigma_{1D}=0.136$ km s$^{-1}$), as shown in Fig \ref{fig:ophse}. These stars are also consistently located high above OphSE's PMS. Applying the age and demographics routine from \citet{Kerr25b}, we find a best fit PARSEC isochronal age of $2.1 \pm  1.4$ Myr, and summarize its properties in Table \ref{tab:micclus}. This population, hereafter OphSE-B, is therefore $\sim$11 Myr younger than OphSE's main component. 

Since \added{OphSE-B is embedded within the extended halo of OphSE, we can test how likely an overdensity like OphSE-B is to emerge by chance. To establish a mean OphSE halo density (including OphSE-B), we compute $d_3$ (distance to the third nearest neighbour) for all stars with $P_{fin}>0.5$ in  $(d_l, d_b, 6v_{T,l}, 6v_{T,b})$ position-velocity coordinates, where $d_l$ and $d_b$ are the on-sky angular separation relative to the average, converted to distance at the average for the population, and the transverse velocities are computed relative to the average distance. This metric avoids \textit{Gaia}'s large distance uncertainties, which are an order of magnitude larger than OphSE-B's sub-pc on-sky size. We exclude stars with $d_3<4$~pc outside OphSE-B to exclude the OphSE's core, where any population at the density of OphSE-B would not be marked as distinct. We set the volume occupied by the star in 4D position-transverse velocity anomaly space to $d_3^4$/3, with the total volume being the sum of that across the N remaining stars.  We then randomly generate N positions within that volume. We find that only 1.8\% of samples reproduce a set of 4 stars as dense as OphSE-B. Even if we assume that this probability exists equally for all 16 populations, an overdensity at the level of OphSE-B in any population is 25\%. For our age false positive tests, we take $10^5$ 3-star samples to match the number of valid PMS stars, and find that only 2 samples produce an age below OphSE-B's upper age limit. The age and clustering false positive probabilities therefore establish OphSE-B as a highly robust structure.}

Like AndS-Cen, OphSE-B's mass is similar to typical multiple star systems, but none of its high-confidence members are flagged as components of resolved binaries in \citetalias{Kerr25b}, and its virial ratio is inconsistent with boundedness, so it is not a multiple star system. The stars surrounding OphSE-B's core have similar velocities, but we find their photometry is consistent with the main component of OphSE.



\subsection{Taurus-Orion 1B} \label{sec:tor1b_ss}

TOR1B has two visibly distinct overdensities in spatial coordinates that we label as components A and B (see Fig. \ref{fig:ophse}). Their best-fit ages differ by $\sim7$ Myr, suggesting that they are distinct in age. \added{However, much of this difference is dependent on the rejection of outliers in TOR1B-A, as without sigma clipping, stars below its tight sequence at 4.5 Myr begin to dominate, and 1-$\sigma$ upper uncertainty interval increases to 11.5 Myr, consistent with component B. Only two of the outliers in TOR1B-A have positions on the CMD consistent with TOR1B-B's sequence, so it is possible that many or all low outliers in TOR1B-A are unrelated to either component, but without deeper RV coverage, it will not be possible to confirm their age distinctness. We therefore classify these substructures as preliminary, requiring membership refinement. If the age outliers of TOR1B-A are confirmed as members with spectroscopic and RV follow-up, it would suggest more internal variation than bulk differences. The bulk velocities of the components are nonetheless within 1 km s$^{-1}$ of each other, so they can be reliably treated as a single unit during traceback. }

\section{Origins of Low-Mass Populations} \label{app:conn_supldisc}

\subsection{Orion and VulE} \label{app:vule}

VulE is the oldest low-mass population that we connect to Orion, with a $30.7 \pm 3.9$ Myr age consistent with NGC 2232 and LP2439 in Monoceros Southwest \citep{Kerr21}. The similar ages and 14 km s$^{-1}$ velocity difference relative to LP 2439 make it unlikely that populations in MSW could have triggered VulE's formation, but a cloud-cloud collision is plausible. The 14 km s$^{-1}$ relative velocity oriented primarily in the galactic plane is near the peak of the collision velocity PDF from \citep{Li18}, indicating that this speed is consistent with a collision driven by galactic shear. This hypothesis could explain the initial triggering of star formation in Orion, but with only one low-mass representative of the potential cloud impacting Orion, this is difficult to verify. 

\subsection{Vela - Cr135} \label{app:vela}

The co-moving or co-spatial populations for AndS and Theia 78 largely define the Collinder 135 (Cr135) family in \citet{Swiggum24}, which comprises much of the Vela Complex \citep[e.g.,][]{CantatGaudin19}. However, compared to the Local and Orion Families, the connection between Cr135 and these low-mass populations is less clear. We show their trajectories in Figure \ref{fig:othertraces}, revealing broad expansion among populations near the bottom of that figure, similar to what was shown by \citep{Swiggum24}. IC 348 is an outlier, and is likely not related, since it is nearly 40 Myr younger than the AndS association that is co-moving with its traced-back progenitor cloud at formation. 

AndS forms with inbound velocities relative to the Cr135 Family's progenitor clouds in Figure \ref{fig:othertraces}. However, unlike VulE, which may have connections to early star formation in Orion, the best fit age for AndS pre-dates the oldest populations in the Cr135 Family, and its close approach does not pass within 60 pc of the progenitor clouds to any populations in the family. Its formation is therefore likely unrelated. AndS is located near the edge of the volume that our literature sample covers when it forms, so even if it does relate to a larger family, that family is likely centered outside our search volume. 

Theia 78 forms near its closest approach to the Cr135 family, although its low Galactic $Z$ at formation places all components outside the 100 pc co-spatial radius. However, its velocities are within 4 km s$^{-1}$ of LP 2383, which \citet{Swiggum24} lists as part of the Cr135 Family. This combination of similar velocities and formation near its closest approach makes some connection between Theia 78 and the Cr135 Family plausible, such as triggered star formation through an interaction with the Family's feedback-driven bubble or an outlying part of its parent cloud. However, Theia 78's largely tangential velocities relative to the Cr135 Family at formation make it unlikely that its formation relates solely to swept-up bubble material. 

\subsection{Ophiuchus Southeast}

The online-only version of Figure \ref{fig:bigtraceback} shows that OphSE is broadly diverging from several populations that \citet{Swiggum24} associates with the M6 Family, such as IC 2391, Alessi 5, and NGC 2451A. However, the association is more than 200 pc from the center of the family at formation, so it does not meet any of our relatedness conditions. It is also more than 15 Myr younger than NGC 3228, the youngest population in our sample located in that family \citep{CantatGaudin20}. Therefore, while a connection between OphSE and the M6 family is possible, our results are inconclusive. 

\subsection{Other Populations} \label{app:otherpops}

CasE and SCYA-54 were co-moving or co-spatial with several populations at formation. However, as indicated in Figure \ref{fig:othertraces}, none of their trajectories indicates a direct relationship. CasE forms co-moving with Alessi 20, ASCC 127, and CFN-1, but does not form within 100 pc of them, and the lack of divergence makes it inconsistent with a common formation event. SCYA-54's subcomponents were both within 1 km s$^{-1}$ of NGC 7058 in Cupavo at formation. However, those populations were separated by more than 140 pc at formation, with a $>20$ Myr age difference between SCYA-54's $29.8 \pm 1.9$ bulk isochronal age and NGC 7058 at 55.3 $\pm$ 2.3 Myr \citep{Kerr24, Kerr25b}. The $\alpha$ Persei cluster is both co-moving and co-spatial with SCYA-54, but its much older $\sim$80 Myr age \citep{GalindoGuil22} makes a common origin even less likely. Platais 8 is co-spatial with SCYA-54-0 at formation, although they are 71 pc apart at formation and their velocities differ by 11 km s$^{-1}$. CasE and SCYA-54 therefore both lack clear common origins with a literature association, although broader connections are plausible, such as an origin in a common spiral arm-scale structure like the Radcliffe Wave \citep{Alves20}. 

The last population, TOR1A, lacks any co-spatial or co-moving populations at formation. This may be due to its inbound velocities relative to the sun, which trace back beyond the volume where our literature sample has coverage (see Fig. \ref{fig:bigtraceback}). Our literature sample may therefore be too volume-limited to assess its origin.

\section{Velocity Asymmetries in CaNMoS and AquENS} \label{app:localvels}

\begin{figure*}
\centering
    \includegraphics[width=16cm]{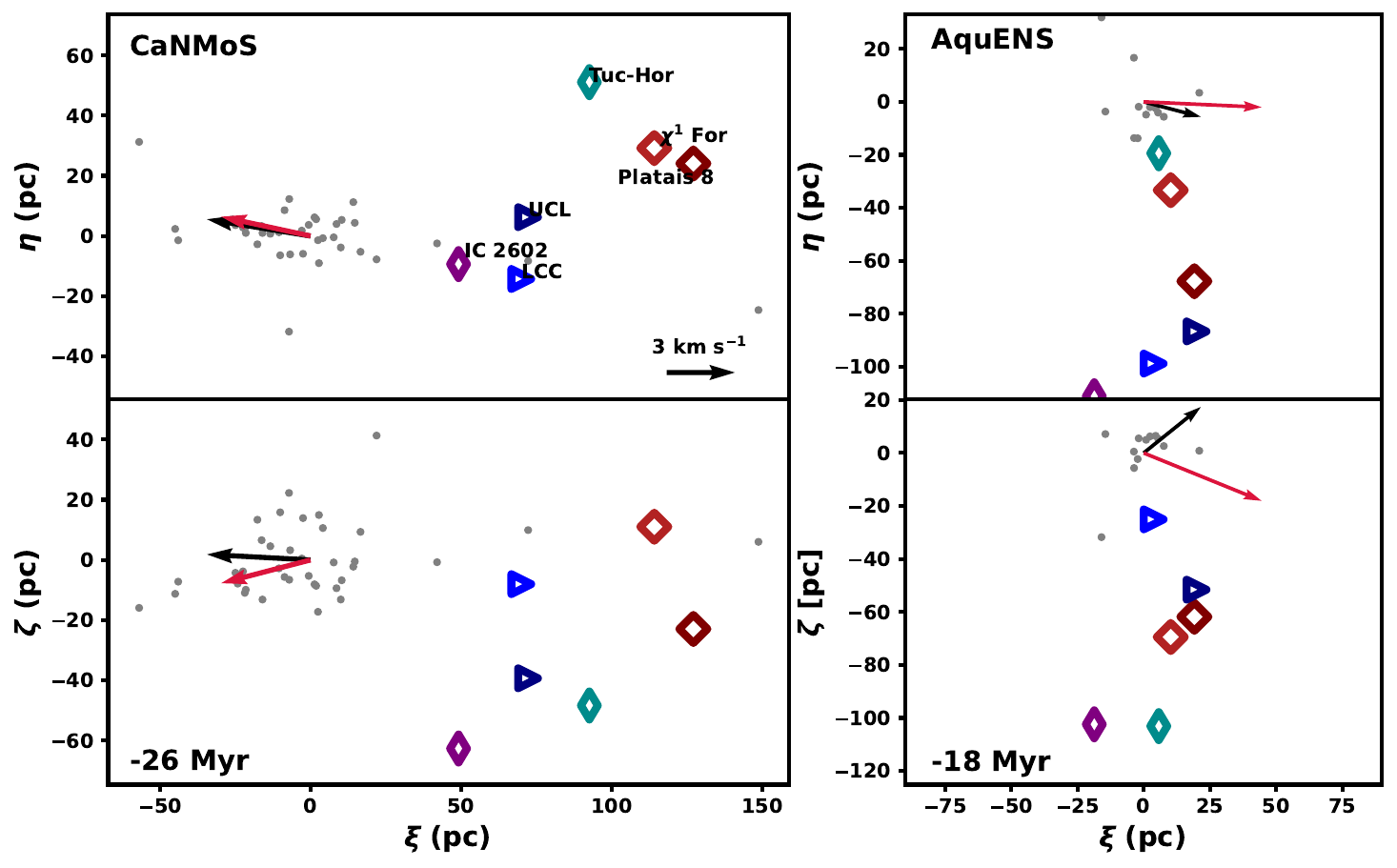}
\caption{Positions of CaMNoS (left) and AquENS (right) members (grey) relative to their average at formation 26 and 18 Myr ago, respectively, plotted against early-forming components of the Local Family (labelled in the first panel). We plot two velocity vectors originating from the average position of each low-mass association, with the black one corresponding to direction of motion, measured against the average of the Local Bubble, and the velocity vector between the most widely-separated subcomponent: Theia 72-1 and CMaN for CaNMoS, and AqE and ScuN-AE for AquENS. CaNMoS shows a velocity difference pointed away from the early components of the Local Family, consistent with a collision between material ejected from those populations and some interloping cloud. In AquENS, the velocity difference is oriented along the direction of motion in all directions but Z, suggesting that ScuN-AE inherited an outward velocity vector through an interaction with the local bubble. } 
\label{fig:veldisps}
\end{figure*}

CaNMoS and AquENS both exhibit highly asymmetric space-velocity distributions that may inform their origins. The major axis of CaNMoS's velocity distribution is more than 5 times longer than its minor axis, resulting in a present day distribution that spans 118 pc and 5.5 km s$^{-1}$ between the centers of CMaN and Theia 72-1. Theia 72's velocities fall within only 1.5 km s$^{-1}$ of $\chi^1$ For, while CMaN has velocities closer to the Local Family's average, suggesting that they may relate to different gas components. The left column of Fig. \ref{fig:veldisps} shows the location of the older populations in the Local Family relative to CaNMoS alongside two velocities: CaNMoS's motion vector relative to the Local Family's average at formation, and the velocity difference between CMaN and Theia 72-1. Both velocities are oriented away from the Local Family's center, so their motion and dispersion align with most stellar feedback. 

The similarity between the velocities of Theia 72 and $\chi^1$ suggests that CaNMoS may have formed in a cloud-cloud collision, most likely between material related to $\chi^1$ For and slower material in the Local Bubble, such as gas from Sco-Cen's progenitor cloud. The orientation of CaNMoS's velocity dispersion along the separation vector to probable feedback sources suggests that an origin in accelerating bubble material is also possible. However, neither CaNMoS component has velocities consistent with our bubble model in Section \ref{sec:bubble-local}, and that model places CaNMoS outside of the bubble at formation. This would require events that cause fast expansion to align CaNMoS with the bubble edge, deceleration to match CMaN's velocity, and quick acceleration to produce Theia 72's 7.1 km s$^{-1}$ outward velocity. This is possible, but the cloud collision scenario provides a simpler explanation. 

We show the velocity difference between AqE and ScuN-AE in AquENS at formation in the right panel of Figure \ref{fig:veldisps}, excluding stars with $v_r<-8.9$ km s$^{-1}$ to remove populations offset from the center of AquENS in the $+\xi$ direction at formation (see Fig. \ref{fig:combpops}). The velocity difference between the components is largely aligned with the direction of motion in the plane of the Galaxy ($\xi$ vs $\eta$), but near-perpendicular to it in $\xi$ vs $\zeta$ space. ScuN-AE therefore hosts velocities that are both faster than AqE in the direction of motion and more radial relative to the local family in the vertical $\zeta$ axis. This vertical velocity change may indicate acceleration from supernovae in the Local Bubble while the progenitor cloud passed, while the component in the plane of the galaxy is a signature common of shear-driven cloud-cloud collisions \citep{Li18}.

\bibliography{rkerr_citations}{}
\bibliographystyle{aasjournalv7}


\end{CJK*}

\end{document}